\catcode`\@=11

\newif\if@fewtab\@fewtabtrue
{\count255=\time\divide\count255 by 60
\xdef\hourmin{\number\count255}
\multiply\count255 by-60\advance\count255 by\time
\xdef\hourmin{\hourmin:\ifnum\count255<10 0\fi\the\count255}}
\def\ps@draft{\let\@mkboth\@gobbletwo
    \def\@oddhead{}
    \def\@oddfoot{\hbox to 7 cm{\tiny \versionno
       \hfil}\hskip -7cm\hfil\rm\thepage \hfil {\tiny\draftdate}}
    \def\@evenhead{}\let\@evenfoot\@oddfoot}
\def\draftdate{\number\month/\number\day/\number\year\ \ \ \hourmin }

\global\def\draftcontrol{0}

\def\draftcite#1{\ifnum\draftcontrol=1#1\else{}\fi}
\def\@lbibitem[#1]#2{\item{}\hskip -3\hbox to 2cm
{\hfil$\scriptstyle\draftcite{#2}$}\hskip
1cm[\@biblabel{#1}]\if@filesw
     {\def\protect##1{\string ##1\space}\immediate
      \write\@auxout{\string\bibcite{#2}{#1}}}\fi\ignorespaces}

\def\@bibitem#1{\item\hskip -3cm \hbox to 2cm
{\hfil {\footnotesize\draftcite{#1}}}\hskip 1cm
\if@filesw \immediate\write\@auxout
       {\string\bibcite{#1}{\the\value{\@listctr}}}\fi\ignorespaces}

\def\citen#1{\if@filesw \immediate\write \@auxout {\string\citation{#1}}\fi%
\@tempcntb\m@ne \let\@h@ld\relax \def\@citea{}%
\@for \@citeb:=#1\do {\@ifundefined {b@\@citeb}%
    {\@h@ld\@citea\@tempcntb\m@ne{\bf ?}%
    \@warning {Citation `\@citeb ' on page \thepage \space undefined}}%
    {\@tempcnta\@tempcntb \advance\@tempcnta\@ne
    \setbox\z@\hbox\bgroup\ifcat0\csname b@\@citeb \endcsname \relax
    \egroup \@tempcntb\number\csname b@\@citeb \endcsname \relax
    \else \egroup \@tempcntb\m@ne \fi \ifnum\@tempcnta=\@tempcntb
    \ifx\@h@ld\relax \edef \@h@ld{\@citea\csname b@\@citeb\endcsname}%
    \else \edef\@h@ld{\hbox{--}\penalty\@highpenalty
    \csname b@\@citeb\endcsname}\fi
    \else \@h@ld\@citea\csname b@\@citeb \endcsname \let\@h@ld\relax \fi}%
\def\@citea{,\penalty\@highpenalty\hskip.13em plus.13em minus.13em}}\@h@ld}
\def\@citex[#1]#2{\@cite{\citen{#2}}{#1}}%
\def\@cite#1#2{\leavevmode\unskip\ifnum\lastpenalty=\z@\penalty\@highpenalty\fi%
  \ [{\multiply\@highpenalty 3 #1%
  \if@tempswa,\penalty\@highpenalty\ #2\fi}]}   %
\makeatother 

\catcode`\@=12

\def\ee           {\mathbf{e}}
\renewcommand\gg  {\mathbf{g}}

\def\nn           {\mathbf{n}}
\def\qq           {\mathbf{q}}
\def\rr           {\mathbf{r}}
\renewcommand\ss  {\mathbf{s}}

\def\KK           {\mathbf{K}}
\def\QQ           {\mathbf{Q}}

\def\al           {\alpha}
\def\be           {\beta}
\def\ga           {\gamma}
\def\de           {\delta}
\def\ep           {\epsilon}
\def\vep          {\varepsilon}
\def\vph          {\varphi}
\def\la           {\lambda}
\def\om           {\omega}
\def\si           {\sigma}

\def\ups          {\upsilon}

\def\GA           {\Gamma}
\def\DE           {\Delta}
\def\LA           {\Lambda}
\def\OM           {\Omega}
\def\PI           {\Pi}

\def\PSI          {\Psi}
\def\SI           {\Sigma}
\def\TH           {\Theta}

\def\cala  {{\cal A}}

\def\calc  {{\cal C}}

\def\calg  {{\cal G}}
\def\calh  {{\cal H}}

\def\caln  {{\cal N}}
\def\calo  {{\cal O}}

\def\dl            {\bf}

\def\naturals      {{\dl N}}

\def\reals         {{\dl R}}
\def\zet           {{\dl Z}}

\def\cft           {conformal field theory}

\def\cfts          {conformal field theories}

\def\cal           {current algebra}

\def\iqhf          {incompressible quantum Hall fluid}
\def\iqhfs         {incompressible quantum Hall fluids}

\def\rep           {representation}

\def\tft           {topological field theory}
\def\tfts          {topological field theories}
\def\twodim        {two-dimensional}

\def\va            {Virasoro algebra}

\def\sutw          {su(2)}

\def\uone          {\hat{u}(1)}
\def\sutwo         {\hat{su}(2)}

\long\def\query#1{\hskip 0pt{\vadjust{\everypar={}\small\vtop to 0pt{\hbox{}%
     \vskip -13pt\rlap{\hbox to 49.0pc{\hfil{\vtop{\hsize=8pc\tolerance=6000%
     \hfuzz=.5pc\rightskip=0pt plus 3em\noindent#1}}}}\vss}}}}%

\newcommand\erf[1] {(\ref{#1})}

\def\futnote#1     {\footnote{~#1}\ }
\def\g             {{\bf g}}

\DeclareMathSymbol{\hodge}{\mathord}{symbols}{"03}
\def\hy            {$\mbox{-\hspace{-.66 mm}-}$}

\def\ii            {{\rm i}}

\long\def\labl#1   {\label{#1}\ee \ifnum\draftcontrol=1
                   \mbox{ }\\[-12 mm]\query{#1}\\[5 mm] \fi}
\long\def\Labl#1#2 {\label{#1#2}\ee\ifnum\draftcontrol=1
                   \mbox{ }\\[-12 mm]\query{#1#2}\\[5 mm] \fi}

\def\tr            {{\rm tr} \, }

\def\wzwts         {WZW\hy theories}

\def\Ad         {{\rm Ad}}

\def\LAE        {\Lambda_e}
\def\LAM        {\Lambda_m}
\def\vir        {\mathrm{Vir}}
\def\nel        {\nu_e}
\def\nfra       {\nu_{\mathrm{frac}}}
\def\mod        {\;\mathrm{mod}\;}
\def\tr         {\mathrm{tr}}
\newcommand{\ord}[1]        {\mathrm{ord}(#1)}

\documentclass[12pt]{article}
\usepackage{amssymb,amsfonts,calc,psfig,epsfig,psfrag,latexsym,wasysym}
\usepackage[intlimits]{amsmath}
\usepackage{amsthm,ifthen}
\begin{document}


\begin{flushright}  {~} \\[-1cm]
{\sf cond-mat/0002330}\\
{\sf ETH-TH/00-3}\\
{\sf PAR-LPTHE 00-07}\\[1 mm]
{\sf February 2000} \end{flushright}

\begin{center} \vskip 15mm
{\Large\bf Universality in Quantum Hall Systems:\\[5mm]
Coset Construction of Incompressible States}\\[22mm]
{\large J\"urg Fr\"ohlich$\;^1$ }, \,
{\large Bill Pedrini$\;^1$ } , \, \\[5mm]
{\large Christoph Schweigert$\;^2$, and}  \,
{\large Johannes Walcher$\;^{1,3}$ }\\[7mm]
$^1\;$ Institut f\"ur Theoretische Physik \\
ETH H\"onggerberg\\ CH\,--\,8093\, Z\"urich\\[5mm]
$^2\;$ LPTHE, Universit\'e Paris VI\\
4 place Jussieu\\ F\,--\,75\,252\, Paris\, Cedex 05\\[5mm]
$^3\;$ Theory Division, CERN\\CH\,--\,1211\, Gen\`eve 23
\end{center}
\vskip 15mm

\begin{abstract}
Incompressible Quantum Hall fluids (QHF's) can be described in the scaling limit
by three-dimensional topological field theories. Thanks to the correspondence 
between three-dimensional topological field theories and two dimensional {\em
  chiral} conformal field theories (CCFT's), we propose to study QHF's from
the point of view of  CCFT's.

We derive consistency conditions and stability criteria for those CCFT's that 
can be expected to describe a QHF. A general algorithm is presented which
uses simple currents to construct interesting examples of such CCFT's. 
It generalizes the
description of QHF's in terms of Quantum Hall lattices. Explicit examples, based
on the coset construction, provide candidates for the description
of Quantum Hall fluids with Hall conductivity 
$\sigma_H=\frac{1}{2}\frac{e^2}{h},\frac{1}{4}\frac{e^2}{h},
 \frac{3}{5}\frac{e^2}{h},\frac{e^2}{h},\ldots$.
\end{abstract}
\newpage


\section{Introduction}
In this paper we reconsider the fractional quantum Hall effect,
using methods from conformal and \tft. Besides improving the foundations of a
theoretical description of quantum Hall fluids in a dissipation-free
(incompressible) state (``incompressible quantum Hall fluid'') in terms of
two-dimensional chiral conformal field theory, our main interest is in showing 
that such a description reproduces many important features of incompressible
quantum Hall fluids corresponding to Hall conductivities
\begin{equation}
  \si_H\;=\;\frac{1}{2}\frac{e^2}{h}\; ,
          \;\frac{1}{4}\frac{e^2}{h}\; ,
          \;\frac{e^2}{h}\; ,\;\ldots
  \quad.
\end{equation}

The quantum Hall effect is observed in two-dimensional ($2D$) electron gases
forming at an interface between a semi-conductor and an insulator when such
gases are put into a strong, uniform magnetic field transversal to the plane
to which the electrons are confined (by an electric field). Tuning the current 
through the sample to a fixed value $I=(I_x,I_y)$\ and measuring the voltage
drops, $(V_x,V_y)$, in the $x$- and $y$-direction, one can determine the
longitudinal and Hall resistances from the \textit{Ohm-Hall law}
\begin{equation}
  \label{eq:ohm-hall}
  \begin{array}{rcl}
    V_x & = & R_LI_x+R_HI_y \\
    V_y & = & -R_HI_x+R_LI_y 
  \end{array}
  \quad.
\end{equation}
For a fixed external magnetic field, one can vary the density of electrons in
the $2D$\ electron gas by varying the gate voltage, i.e., the electric field
perpendicular to the interface to which the electron gas is confined. One then 
finds that if the electron density belongs to certain intervals (whose width
depends on the magnetic field and the strength and density of impurities) the
longitudinal resistance vanishes: $R_L=0$. This is a signal for the
\textit{absence of dissipative processes} in the $2D$\ electron gas, which, in 
turn, can be interpreted as an indication that the gas is in an
``\textit{incompressible state}'' with a strictly positive mobility gap in the
bulk of the sample. Experimentally, one finds that whenever $R_L$\ vanishes
the Hall conductivity
\begin{equation}
  \si_H=R_H^{-1},\quad\mathrm{for}\ R_L=0,
\end{equation}
is a \textit{rational multiple} of $\frac{e^2}{h}$, where $e$\ is the
elementary electric charge and $h$\ is the Planck constant. This
``\textit{quantization}'' of the Hall conductivity is extremely precise for
well developed plateaux. The phenomena described here are referred to as the
\textit{quantum Hall effect}.

The \textit{integer} quantum Hall effect
($\si_H=1\frac{e^2}{h},2\frac{e^2}{h},3\frac{e^2}{h},\ldots$) was discovered 
by von Klitzing, Dorda and Pepper in 1980 \cite{kdp}, the \textit{fractional}
quantum Hall effect by Tsui, St\"ormer and Gossard in 1982
\cite{tsui}. Fundamental insights into theoretical explanations of these
remarkable effects were soon brought forward by Laughlin
\cite{lauint,laufra}. In particular, he discovered a trial wave function 
accurately encoding properties of the ground state of a quantum Hall fluid in
an incompressible state with $\si_H=\frac{1}{3}\frac{e^2}{h}$, which is now 
called
``Laughlin fluid''. He made the tantalizing observation that, in a Laughlin
fluid, there are quasi-particles of electric charge $\pm\frac{e}{3}$, called
``Laughlin vortices''. It was recognized that, besides their fractional electric 
charge, Laughlin vortices carry one quantum of flux and exhibit fractional
(``braid'') statistics; see \cite{asw,gipr} and references given there. In 
1982, Halperin showed that quantum Hall fluids in an incompressible state
confined to a finite domain exhibit chiral diamagnetic currents localized near 
the boundary of the sample, the so called \textit{``edge currents''}
\cite{halp}. Halperin's arguments were based on a direct analysis of the
quantum mechanics of $2D$\ non-interacting electron gases under the
influence of a transverse magnetic field and of impurities in the bulk of the
sample. The important r\^{o}le played by edge currents was emphasized in later 
work by B\"uttiker \cite{butt}, Beenaker \cite{been}, and others
\cite{other1}.

In 1989/90, it was recognized independently by Wen \cite{wen1} and by
Fr\"ohlich and Kerler \cite{frke} (see also \cite{other2} for a sample of
subsequent work) that the diamagnetic edge currents of an arbitrary quantum
Hall fluid in an incompressible state are described by quantum mechanical
current operators generating a \textit{chiral current} (Kac-Moody)
\textit{algebra}. This opened the view towards using methods from (chiral)
conformal field theory to analyze incompressible quantum Hall fluids.

It was emphasized in \cite{frke} that, in order to study \textit{universal}
properties of quantum Hall fluids in an incompressible state, it is convenient 
to describe such fluids in the so called \textit{scaling limit} in which
distances and times are infinitely rescaled. The concept of studying physical
systems in the scaling limit is familiar from the theory of \textit{critical
  phenomena}. It was noticed that, because quantum Hall fluids in an
incompressible state exhibit a strictly positive mobility gap, their bulk
properties are described, in the scaling limit, by a \textit{topological field
  theory}, \cite{frke,frze,wen2}. In particular, the theory
describing the quantum-mechanical electric charge- and current density
operators in the scaling limit was shown to be an abelian topological
\textit{Chern-Simons theory}, \cite{frke,frze}. This observation also
provided additional insight into the origin of the diamagnetic edge currents:
they are necessary to guarantee that the total electric charge is
\textit{conserved} in \textit{closed} incompressible quantum Hall fluids;
(``\textit{anomaly cancellation}'' - it is actually quite fascinating to realize 
that the diamagnetic edge currents of a quantum Hall fluid in an
incompressible state are carried by chiral, quantum-mechanical degrees of
freedom violating electromagnetic gauge invariance; this violation of
electromagnetic gauge invariance is exactly compensated by one exhibited by the 
bulk degrees of freedom of the fluid). The conspiracy between edge and bulk
degrees of freedom in ensuring conservation of the total electric charge and
in cancelling each other's violations of electromagnetic gauge invariance is
an instance of what has recently become known as the ``\textit{holographic
  principle}'', \cite{suss}. In the example of incompressible quantum Hall
fluids this principle would say that the Hall conductivity $\si_H$\ can be
measured in experiments involving only edge currents, or in ones involving
only bulk currents, or in experiments involving edge \textit{and} bulk
currents,  and that there is a
correspondence between the quasi-particle spectra in the bulk and the
quasi-particle spectra of the edge degrees of freedom. In more theoretical
terms, the \textit{topological field theory} describing the scaling limit of
the bulk of an incompressible quantum Hall fluid is completely determined by a 
\textit{chiral conformal field theory} describing the edge degrees of freedom
of an incompressible quantum Hall fluid with the same Hall conductivity. The
connection between three-dimensional topological Chern-Simons theory and the
two-dimensional chiral Wess-Zumino-Witten model (Kac-Moody algebra) was
discovered by Witten \cite{witt27}. A fairly general construction of
$3D$\ topological field theories from $2D$\ chiral conformal field theories
was later described in \cite{frki}. It will be invoked in this paper.

In general, there is no guarantee that the chiral conformal field theory that
determines the topological field theory describing the scaling limit of the
\textit{bulk} of an incompressible quantum Hall fluid is identical to the one
describing the \textit{edge} degrees of freedom of the fluid; although some of 
their properties do coincide. The reason is that the structure of the fluid
near the boundary of the domain to which it is confined can be quite
complicated, exhibiting several distinct, thin layers. Thus, the theory of the 
edge degrees of freedom is, in general, more complicated, and less universal
than the theory of the bulk. This is why we shall emphasize the study of
topological field theories describing the scaling limit of the \textit{bulk}
of \textit{homogeneous} quantum Hall fluids in incompressible states, i.e.,
fluids whose Hall conductivity and quasi-particle spectra are everywhere the
same in the bulk.

There is no doubt that the idea to analyze universal properties of quantum
Hall fluids in incompressible states by considering their scaling limits and
using topological field theories to describe them is sound and has turned out
to be fruitful. The key question is then how much \textit{concrete information} about incompressible quantum Hall fluids can be gained from
such a general and quite abstract approach. Clearly, this approach
\textit{cannot} be used to understand for which values of the external control
parameters, such as the magnetic field, the electron density, the density and
strength of impurities, etc. the ground state of a quantum Hall fluid is
incompressible, in the sense that a mobility gap opens and the longitudinal
resistance $R_L$\ vanishes. An understanding of these problems requires
analysis of the microscopic quantum mechanics of $2D$\ interacting electron
gases in a transverse magnetic field, and this is hard analytical and/or
numerical work; see \cite{gipr,leezh,morf}. However,
\textit{assuming} that $R_L$\ \textit{vanishes}, methods from $3D$\ \tft/$2D$\ 
chiral \cft\ provide a \textit{key to understand} which values the Hall
conductivity can take, what spectra of quasi-particles may occur in
incompressible states and what their quantum numbers are, what fractional
electric charges may be measured, whether there may be \textit{several
  distinct} incompressible states corresponding to the \textit{same} value of
the Hall conductivity (``intra-plateaux transitions''), what kind of heat
currents may be observed, etc.; see
\cite{wen1,frth,fkst,frst}. The main purpose of our paper is
to improve the foundations and generalize the scope of this approach. The
pay-off will be to provide very plausible descriptions of incompressible
quantum Hall fluids with Hall conductivities
$\si_H=\frac{1}{2}\frac{e^2}{h},\frac{1}{4}\frac{e^2}{h},
\frac{e^2}{h}$,\ldots\ 
Accurate descriptions of incompressible quantum Hall fluids with
$\si_H=\frac{N}{2pN+1}\frac{e^2}{h},\ p=1,2,3,\; N=1,2,\ldots,8,\ldots$\ 
have been presented, within the general approach described above, in
\cite{frth,fkst,frst}. They have features fairly closely related
to e.g. Jain's description of these fluids \cite{jain}. But the approach in
\cite{frth,fkst,frst} was not quite general enough to provide a
plausible description of e.g. an incompressible quantum Hall fluid with Hall
conductivity $\si_H=\frac{1}{2}\frac{e^2}{h}$, which is observed in
\textit{double-layer systems}. In the scaling limit, the theoretical
description of such a fluid can be expected to have two important
\textit{symmetries}, an $SU(2)$-layer symmetry and the $SU(2)$\ of quantum
mechanical spin; (see e.g. \cite{frst}). It can happen, however, that the
diagonal $SU(2)$-subgroup of the symmetry group $SU(2)_{\mathrm{layer}}\times
SU(2)_{\mathrm{spin}}$\ is \textit{not} a global symmetry of the fluid - it is 
``gauged''. [Adding an electron with ``spin up'' in one layer may turn out to
be ``gauge-equivalent'' to adding an electron with ``spin down'' in the other
layer - roughly speaking.] This possibility appears to be realized in the
\textit{``Pfaffian state''} proposed by Moore and Read \cite{more,mire}. One
concrete goal of this paper is to generalize the approach of
\cite{frth,fkst,frst} by ``gauging'' subgroups of symmetry groups
of incompressible quantum Hall fluids. 
In several physically interesting examples, this construction, usually 
referred to as the \textit{coset construction} \cite{baha,goko}, does 
\textit{not} change the
value of the Hall conductivity; but it changes the quasi-particle spectrum (by 
identifying certain quasi-particles that were formerly distinct and by
turning some formerly elementary quasi-particles into composite
quasi-particles); see Appendix \ref{app:cosets} for a more precise treatment.
It leads to
a natural theoretical interpretation of various trial ground-state wave
functions, such as the one in \cite{more}. Some examples of our general
approach have been described in \cite{cagt}, but with an emphasis on
properties of edge states rather than of bulk states.

Historically, the problem of the \textit{quantization} of the Hall conductivity
of quantum Hall fluids in incompressible states and of elucidating other
physical properties of such fluids has of course been studied since the discovery of the 
quantum Hall effect. A highly original theory of quantum transport has been
developed, for this and other purposes, by Thouless and coworkers
\cite{thknn} and their followers \cite{avse}; see also \cite{frad} and
references given there. In this approach the Hall conductivity is identified
with a Chern number. An alternative approach, identifying the Hall
conductivity with an index, was proposed in
\cite{belli,avss,aigr}. An approach towards understanding the
quantization of the Hall conductivity in terms of edge states has been
described in \cite{frgw}; see also
\cite{lauint,mamapu,bipu,alcf}. In all these approaches, the 
$2D$\ electron gas is treated as \textit{noninteracting}, which severely
limits their scope. The observation that one can come up with \textit{general
  predictions} of the possible values of the Hall conductivity, of the
spectrum of quasi-particles and of their quantum numbers, such as their
(generally fractional) electric charges, of incompressible quantum Hall fluids 
by merely assuming that the longitudinal resistance $R_L$\ vanishes and then
using methods from \tft/chiral \cft\ was made in
\cite{frke,frze,wen2}. It is based on a fundamental connection between
the electric charge of a cluster of quasi-particles and its quantum
statistics, which was first described in \cite{frke}. Besides an analysis of
concrete examples, a general implementation and improved presentation of these 
ideas is among the main purposes of our paper.

We conclude this introduction with a brief outline of the contents of this
paper. In Sect. \ref{sec:generalfeatures}, we first recall the laws of
electrodynamics of quantum Hall fluids in an incompressible state ($R_L=0$),
starting from the \textit{Ohm-Hall law}. We then show that these laws alone imply
the existence of (``anomalous'') chiral edge currents generating a chiral
current (Kac-Moody) algebra. Subsequently, the connection between the laws of
electrodynamics of incompressible quantum Hall fluids and Chern-Simons theory
and between the latter and the ``anomalous nature'' of the edge currents is
briefly recalled. We then review how a description of incompressible quantum
Hall fluids in the scaling limit leads one to consider $3D$\ topological
field theories generalizing the electromagnetic Chern-Simons theory. We argue
that those topological field theories that are relevant in a theoretical
description of incompressible quantum Hall fluids in the scaling limit can be
constructed from $2D$\ chiral conformal field theories, and we sketch some
basic aspects of this construction; (for details see \cite{frki}). Of course,
chiral conformal field theories also appear in the description of the edge
degrees of freedom of incompressible quantum Hall fluids with the same Hall
conductivity.

In Sect.\ \ref{sec:conditions}, we formulate fundamental conditions,
``\textit{consistency conditions}'', singling out those topological field
theories/chiral conformal field theories that can, in principle, appear in the 
description of the scaling limit of an incompressible quantum Hall fluid. We
comment on the r\^{o}le played by an intriguing mathematical property of such
theories, ``modular covariance'', in determining the full spectrum of
quasi-particles of an incompressible quantum Hall fluid (clarifying, perhaps,
some misconceptions that have appeared in the literature). We then recall some 
phenomenological criteria enabling one to assess the \textit{stability} of an
incompressible quantum Hall fluid described by a given \tft. They represent an 
elaboration on criteria proposed in \cite{frth,fkst,frst}.

In Sect.\ \ref{sec:lattices}, we recall the construction of a \textit{special
  class} of \textit{topological field theories} relevant for a theoretical
description of incompressible quantum Hall fluids that was identified and
studied in \cite{frth,fkst,frst}. Theories in this special class
are in a one-to-one correspondence with certain (odd) integral lattices called 
\textit{``quantum Hall lattices''}. The main properties of quantum Hall
lattices are recalled, and the consistency conditions and stability criteria
formulated in Sect.\ \ref{sec:conditions} are used to derive some constraints
that must be imposed on quantum Hall lattices.

In Sect.\ \ref{sec:recipe}, a more general class of topological field theories/chiral 
conformal field theories expected to be relevant for a description of
incompressible quantum Hall fluids is identified. It is explained how to
calculate the Hall conductivity in such theories and why it is necessarily a
\textit{rational multiple} of $\frac{e^2}{h}$. The value of the smallest
fractional electric charge that can appear as the charge of a quasi-particle
of a quantum Hall fluid described by such a theory is calculated. It is then
indicated how examples of topological field theories from the class of
theories described at the beginning of Sect.\ \ref{sec:recipe} can be
constructed from the theories described in Sect.\ \ref{sec:lattices}, which are
characterized by quantum Hall lattices, by making use of the so called
\textit{coset construction}, \cite{baha,goko,fusS4}.

In Sect. \ref{sec:examples}, we discuss \textit{concrete examples} of \tfts\
describing interesting quantum Hall fluids in incompressible states. These
examples are based on ``Virasoro minimal models'', such as the $2D$\ chiral
Ising model, and ``simple current extensions'' thereof. Among our examples are 
theories describing \iqhfs\ with Hall conductivity
$\si_H=\frac{1}{2}\frac{e^2}{h},\frac{1}{4}\frac{e^2}{h}$\ and
$\frac{e^2}{h}$. Perhaps, the most interesting example is a theory with
$\si_H=\frac{1}{2}\frac{e^2}{h}$, which is related to the $2D$\ chiral Ising
model. It predict properties of an incompressible quantum Hall state at
$\si_H=\frac{1}{2}\frac{e^2}{h}$\ compatible with those predicted by the
``Pfaffian state'' of Moore and Read \cite{more}. Our results on specific
examples of \iqhfs\ are summarized in four tables.

In Appendix \ref{app:modular}, ``modular covariance of the theories described
in Sects.\ \ref{sec:recipe} and \ref{sec:examples} is discussed, and in
Appendix \ref{app:cosets} some basic facts concerning the ``coset
construction'' are recalled.

We dedicate this paper to the memory of Quin Luttinger, who made fundamental
contributions to diverse fields of theoretical physics ranging from
relativistic QED over condensed-matter physics to mathematical physics. With
his open mind, his charming, friendly personality, his curiosity and his
original thinking he inspired colleagues in every field to which he turned his 
interest. One of his far-reaching contributions concerned the theory of
one-dimensional electron gases, systems now known as \textit{Luttinger
  liquids}. The chiral degrees of freedom in an \iqhf\ represent an example of 
a one-dimensional gas of electrons or holes. It is sometimes called a ``chiral 
Luttinger liquid''. It is plausible that Luttinger would have found these
systems interesting.

\textbf{Acknowledgements}. One of us (J.F.) has greatly profitted from
collaborations and/or many discussions with A. Alekseev, V. Cheianov, 
K.\ Ensslin, T. Kerler, R. Morf, U. Studer, E. Thiran, X.G. Wen and A. Zee.

\section{General features of Quantum Hall Fluids}
  \label{sec:generalfeatures}

A quantum Hall fluid (QHF) is a two-dimensional interacting electron gas in a 
compensating uniformly charged background, subject to a magnetic field transversal 
to the confinement plane. Among the experimental control parameters is the
filling factor defined as
\begin{equation}
  \label{eq:fillfact}
  \nu = \frac{n^{(0)}}{B^{(0)}/\frac{hc}{e}}
\end{equation}
where $n^{(0)}$\ is the electron density, $B^{(0)}$\ denotes the uniform
  transverse magnetic field, and $\frac{hc}{e}$\ is the flux quantum.

Electric transport properties of a QHF in a small electric field at low
frequency are described by
the relation between the electric field parallel to the plane of the sample
and the (expectation value of the quantum mechanical) electric current,
\begin{equation}
  \label{eq:conduct}
  \vec{J}=
  \left(
    \begin{array}{cc}
    \si_L & \si_H \\ -\si_H & \si_L
    \end{array}
  \right) \vec{E}
  \quad,
\end{equation}
(the Ohm-Hall law, compare \erf{eq:ohm-hall}), 
where $\si_L$\ is the longitudinal conductivity and $\si_H$\ the transverse or 
Hall conductivity.

Experimentally, it is observed that, in certain intervals of the filling
factor, the longitudinal conductivity vanishes \cite{tsui}, a sign that dissipative
processes are absent in the fluid. 
Moreover, it is observed that, on such intervals, the
Hall conductivity is a rational multiple of $\frac{e^2}{\hbar}$. For reasons
that will become clear later, we call a QHF with these properties
``incompressible''.

\subsection{Electrodynamics of \iqhfs}
The basic equations of the electrodynamics of an incompressible QHF can be
derived as follows (see \cite{frke}): In (2+1) dimensions, the electromagnetic 
field tensor is given by
\begin{equation}
  \label{eq:fieldstrength}
  F_{\mu\nu}=
  \left(
    \begin{array}{ccc}
    0 & E_1 & E_2 \\
    -E_1 & 0 & B \\
    -E_2 & -B  & 0 \\
    \end{array}
  \right)
  \quad,
\end{equation}
where $E_1,E_2$\ are the components of the electric field in the sample plane
and $B$\ is a perturbation of the transverse background magnetic field,
such that $B^{(\mathrm{total})}=B^{(0)}+B$. From the (2+1)-dimensional
homogeneous Maxwell equations (Faraday's induction law),
\begin{equation}
  \label{eq:maxhomo}
  \partial_{\mu}F_{\nu\la}+\partial_{\nu}F_{\la\mu}+\partial_{\la}F_{\mu\nu}=0
  \quad,
\end{equation}
the continuity equation for the electric current density (conservation of
electric charge),
\begin{equation}
  \label{eq:conselchar}
  \partial_{\mu}J^{\mu}=0
  \quad,
\end{equation}
and from the transport equation \erf{eq:conduct} for $\si_L=0$, i.e.,
\begin{equation}
  \label{eq:conducthall}
  \vec{J}=
  \left(
    \begin{array}{cc}
    0 & \si_H \\ -\si_H & 0
    \end{array}
  \right) \vec{E}
  \quad,
\end{equation}
it follows that
\begin{equation}
  \label{eq:densityhall}
  J^0=\si_H\; B \quad.
\end{equation}
Equations \erf{eq:conducthall} and \erf{eq:densityhall} can be combined to
the equation
\begin{equation}
  \label{eq:currentbulk}
  J^{\mu}=\si_H\; \ep^{\mu\nu\la}\; F_{\nu\la} 
  \quad.
\end{equation}
of Chern-Simons electrodynamics \cite{jadete}. It describes
the response of an incompressible QHF to an external electromagnetic field. 
It is compatible with the continuity equation \erf{eq:conselchar}
iff $\si_H$\ is constant.

We may take into account the finite extension of the sample confined
to a region $\OM=D\times\reals$\ of space-time, by setting the Hall
conductivity $\si_H$\ to a nonzero, constant value on $\OM$\ and to zero
outside, i.e., 
\begin{equation}
  \label{eq:sigmaonsample}
  \si_H(\cdot)=\si_H\;\chi_{\OM}(\cdot)
  \quad,
\end{equation}
where $\chi_{\OM}(\cdot)$\ is the characteristic function of the space-time
region $\OM$. The divergence of the electric current density
\erf{eq:currentbulk}  is then different from zero on the boundary of the
sample: 
\begin{equation}
  \label{eq:anomconselchar}
  \partial_{\mu}J^{\mu}=
      \si_H \; \ep^{\mu\nu\la} \; \partial_{\mu}\chi_{\OM} \; F_{\nu\la}
  \quad.
\end{equation}

Since conservation of electric charge in closed systems is a law of nature, 
there must be an electric current $J_{{\rm edge}}$\ 
localized at the boundary $\partial\OM$\ of the sample,
with the property that the \textit{total} electric current
\begin{equation}
  \label{eq:currenttotal}
  J_{\rm{total}}^\mu=J^\mu+J_{{\rm edge}}^\mu
\end{equation}
is divergencefree. The edge current, $J_{{\rm edge}}^\mu$, has the form
$J_{{\rm edge}}^\mu= j^\mu \delta_{\partial\Omega}$, where $j^\mu$
is a current density on the boundary $\partial\Omega$ whose component
normal to $\partial\Omega$ must vanish. Equation (\ref{eq:anomconselchar})
then implies that 
\begin{equation}
  \label{eq:boundaryanomaly}
  \partial_{\al}j^{\al}=\frac{1}{2}\si_H\;\ep^{\al\be}F_{\al\be}\quad.
\end{equation}
Here, the indices $\al,\be$\ refer to coordinates for the (1+1)-dimensional
boundary $\partial\Omega$.

Equation (\ref{eq:boundaryanomaly}) expresses the (1+1)-dimensional (abelian) 
chiral anomaly, see e.g.\ \cite{Jac}, 
and tells us that there are chiral (and hence
gapless) degrees of freedom localized on the boundary, which are coupled to
the electromagnetic gauge field in such a way that the electric current they
carry obeys the anomaly equation \erf{eq:boundaryanomaly}. 
Quantum mechanically, this current is described by
a $\uone$-current algebra, with an anomalous commutator proportional to the
Hall conductivity:
\begin{equation}
  \label{eq:u1currentalg}
  [j_m,j_n]\;=\;\de_{m+n,0}\;\si_H\quad ,
\end{equation}
where $j_n$ is the $n^{{\rm th}}$ Fourier component of $j$; see
\cite{frke,wen1,frst}.

\subsection{Topological Field Theory and \iqhfs}
Next, we return to studying the physics of the bulk of the sample. 
The absence
of dissipation in the transport of electric charge \erf{eq:conducthall} can be 
explained by the existence of a gap in the energy spectrum between the ground
state energy of the QHF and the energies of excited (extended) bulk states \cite{frke,wen1}. 
This explains 
the term ``incompressible'': it is not possible to add an additional electron
to the fluid, or to extract one from the fluid,  by paying only an arbitrarily 
small energy. An important consequence of incompressibility is that the
total electric charge is a good quantum number to label different sectors
of physical states at zero temperature.

We are interested in the physics in the scaling limit of an incompressible
QHF, i.e., in the limit in which short-distance- and high-frequency properties 
become invisible. This limit is defined as
follows \cite{frke,frst}. For an arbitrary disc $D$, consider
the family of fluids confined to discs of different sizes
$D_{\TH}=\{x|\TH^{-1}x\in D\}$. The Green functions in the scaling limit can
be constructed from the Green functions of the systems confined to
$D_{\TH}$ as follows:
\begin{equation}
  G^D_{\la_1,\ldots,\la_n}(x_1,\ldots,x_n)\;=\;
  \lim_{\TH\rightarrow\infty}\TH^{\si}
  \langle T[\phi_{\la_1}(\Theta x_1)\ldots\phi_{\la_n}(\Theta x_n)]
  \rangle_{(D_\TH)}
  \quad,
\end{equation}
where the $\phi_{\la}$\ are fields of the theory that describes the
fluid, and the exponent $\si(\la_1,\ldots,\la_n)$\ on the right hand side takes 
into account the scaling dimensions of the fields appearing in the 
time-ordered product. Thus, in the scaling limit, the fluid is considered to 
be confined to a standard disc $D$. Our goal is to describe the space of 
physical state vectors of an incompressible QHF in the scaling limit.

The presence of a positive energy gap
implies that, in this limit, the theory describing an incompressible QHF is a
``\tft'', whose excitations are \textit{static pointlike sources}
localized in the bulk and labelled by quantum numbers, such as electric
charge, or, perhaps, spin. More formally, the pointlike sources are marked 
by elements $\la$\ in a set $\LA$, characteristic of the fluid, which
generates a {\em fusion ring}, a term coming from representation theory
that will be defined below.

Equation \erf{eq:currentbulk}, which relates the ground state expectation value
of the electric current to the external electromagnetic field, is an
expression of the fact that the bulk theory \textit{in the scaling limit} is
\textit{topological}. 
To see this, consider the 
topological (metric-independent) (2+1)-dimensional Chern-Simons (CS) action
\begin{equation}
  \label{eq:csa}
  CS_3[A]=\int d^3x \; \si_H\; \ep^{\mu\nu\la} A_{\mu}\partial_{\nu}A_{\la}
  \quad.
\end{equation}
This is the effective action of the bulk degrees of freedom in an external
electromagnetic field with vector potential $A_\mu$. In fact, it gives the 
correct equation for the electric current density:
\begin{equation}
  \label{eq:currentfromcsa}
  J^{\mu}(x)\;=\;\frac{\de}{\de A_{\mu}(x)}CS_3[A]\;=\;
             \si_H\; \ep^{\mu\nu\la}\; F_{\nu\la} \quad.
\end{equation}
For the current-current correlation function, the CS effective
action yields
\begin{equation}
  \label{eq:current-current}
  \langle J^{\mu}(x) J^{\nu}(y) \rangle\;=\;
    \frac{\de^2}{\de A_{\mu}(x)\de A_{\nu}(y)}CS_3[A]\;=\;
    \si_H \ep^{\mu\nu\la} \partial_{\la}\de(x-y)
  \quad,
\end{equation}
which is the unique expression for the leading term in the scaling limit for a 
system with broken parity, as can be deduced from dimensional analysis \cite{frst}.

The CS action is not invariant under gauge transformations that do not vanish
on the boundary $\partial\OM$\ of the sample space-time. But its gauge
variation is
exactly compensated by the variation of the effective action for the coupling
of the electromagnetic gauge field to the boundary degrees of freedom
described by the $\uone$-current algebra, i.e., the \twodim\ anomalous chiral
action, \cite{Jac}.

\subsection{Two-dimensional chiral \cft \\ and three-dimensional \tft}
In the last subsections we have argued that the Green functions of the
electromagnetic current density of an incompressible QHF are described, in the 
scaling limit, by a ``\tft''; see
\erf{eq:currentfromcsa} and \erf{eq:current-current}.
More generally, the
entire physics of an incompressible QHF can be encoded, in the scaling limit,
into a ``\tft'' defined over the three-dimensional space-time of the sample. 
The purpose of this section is to recall what one means by a ``three-dimensional
topological field theory'', and to explain the connection between some of
these theories and ``two-dimensional chiral conformal field theories''. 

Generally speaking, a \textit{three-dimensional \tft\ } ($3D$\ TFT) associates a
topological invariant to every three-dimensional manifold without boundary. 
However, a finite Hall-sample has a boundary. 
Its space-time is therefore described by a three-dimensional
manifold \textit{with} boundary.
Thus, it must be possible to define those $3D$\ TFT's that describe the physics
of incompressible QHF's in the scaling limit on three-dimensional manifolds
with boundary, provided the boundary is given a suitable geometric
structure. 
Furthermore, it is reasonable to expect that some kind of
"holographic principle" is valid: the TFT on a three-dimensional manifold,
$\OM$, with boundary $\partial\Omega$, denoted by $\SI$, should be unambiguously 
determined by a two-dimensional field theory defined over $\SI$.

\begin{figure}[!tb]
\begin{center}
\input{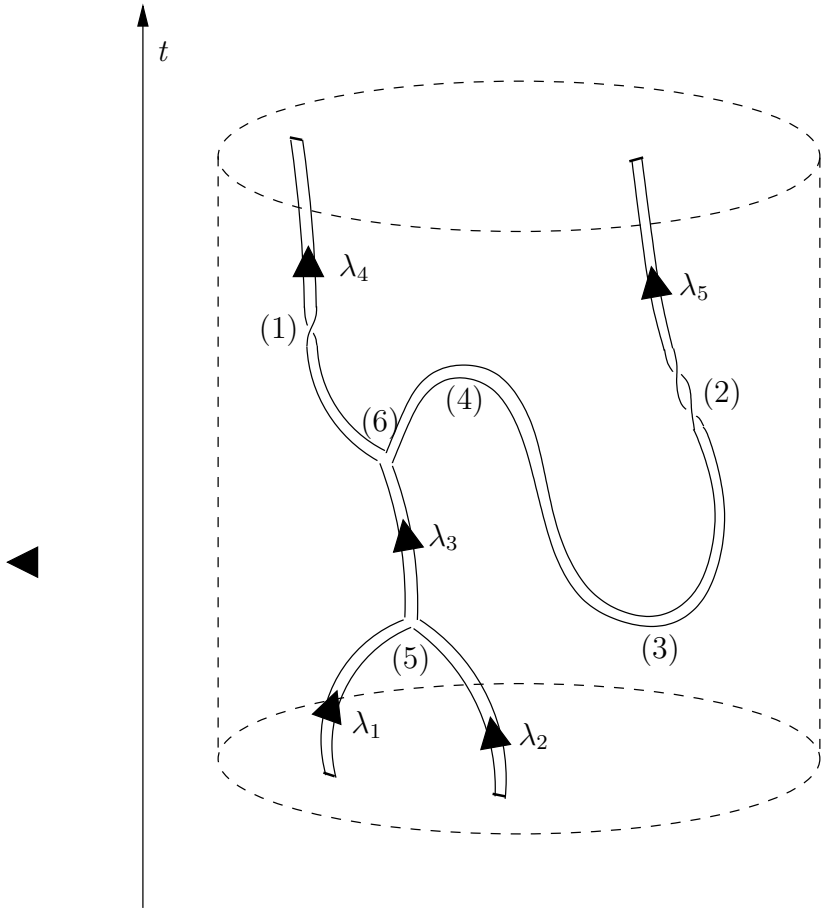}
  \parbox{14cm}{\caption{A Feynman diagram or ribbon graph for some
      sources. One can recognize (1) a self-twist, (2) a double self-twist,
      (3) a pair production, (4) a pair annihilation, (5) two sources that
  \label{fig:ribbon}
      fuse into one, and (6) a source that splits in two.}}
\end{center}
\end{figure}

In the bulk of an incompressible QHF there may be static sources labelled by some
quantum numbers in a fusion ring. 
In rescaled space-time, such sources trace out worldlines. 
At certain times, two sources may collide,
i.e., their worldlines may be fused into a single worldline, or a source may
split into two distinct sources.
Thus, the worldlines of sources in the bulk of an incompressible QHF can be
viewed as the lines of "Feynman diagrams" with trivalent 
vertices\footnote{Higher vertices can be reduced to trivalent vertices by 
repeated fusing.},
each oriented line in the
diagram carrying the quantum numbers of the source it represents; see Figure
\ref{fig:ribbon}.
Since the effective field theory describing an incompressible QHF in the
scaling limit is topological, it is only the {\em topology} of the Feynman
diagram representing the worldlines of the sources that matters.
It is important to realize that the Feynman diagrams are "framed"
diagrams, i.e., along each line in the diagram a field of vectors 
(perpendicular to the tangent vector at each point of the line) is defined which
enables us to keep track of the "self-twist" of the line. 
Such "self-twists" appear as Aharonov-Bohm type phase factors
in the quantum mechanical transition amplitudes of an incompressible QHF.
In fact, the value of a Feynman diagram representing the worldlines of bulk
sources in an incompressible QHF is nothing but a ``\textit{generalized
  Aharonov-Bohm phase}'' depending only on the \textit{topology} of the
diagram, including the \textit{self-twists} of its lines, 
i.e., depending only on the ``\textit{ribbon graph}''
traced out by the Feynman diagram.

Thus, in order to describe incompressible QHF's in the scaling limit, we are
looking for $3D$\ TFT's which can be defined on manifolds with
boundary in which some oriented ribbon graph is inscribed, whose lines are
decorated with quantum numbers from a fusion ring. Moreover, these TFT's 
should be determined by
field theories on the {\em boundary} of the three manifold in such a way that the
"holographic principle" is satisfied.

It turns out that every two-dimensional chiral \cft\ ($2D$\ CCFT)
determines a $3D$\ TFT with all the properties required above. In fact, it is
reasonable to expect that every $3D$\ TFT with the above
properties can be derived from a $2D$\ CCFT.
We assume this as a justification to constrain the class of $3D$\ TFT's used
to describe the physics in the bulk of an \iqhf\ in the scaling limit to those 
derived from $2D$\ CCFT's.
[These CCFT's are, however, not unique: the \wzwts\ based on $so(n)$ and
$so(n+16)$ at level one, e.g., provide identical $3D$\ TFT's.] 
Thus, we recall what is meant by a two-dimensional chiral \cft.

A \textit{two-dimensional chiral \cft\ } is a quantum field theory 
defined over a cylindrical space-time $\reals\times S^1_R$ of radius $R$, 
with coordinates
$(t,\vph)$. The quantum-mechanical degrees of freedom are \textit{chiral},
which means that the dynamical modes of such a theory are purely left- or
purely right-moving. 
\footnote{The choice of left moving modes or right moving modes endows
the boundary with an orientation. In fact, CCFT can be naturally considered
also on closed, oriented surfaces of higher genus. In the description
of a QHF, such surfaces do not appear in physically meaningful situations.}
Put differently, all the fields of a $2D$\ CCFT only depend on \textit{one}
light-cone coordinate, say $u_-=\ups t-R\vph$, with $\ups$\ the propagation
velocity of the modes. 
Among these fields, we consider all the \textit{local}
ones. With the help of the operator product expansion one shows that the
local chiral fields form an algebra of operators (operator-valued
distributions) on the Hilbert-space of physical states.
This algebra is called the \textit{chiral algebra} of the theory and is denoted by
$\cala$. 
Among the fields generating $\cala$, there is the energy-momentum
tensor of the theory. Because the theory is assumed to be chiral, its
energy-momentum tensor is \textit{traceless}. Its Fourier components (with
respect to the coordinate $\vph$), $L_n$, then satisfy the
commutation relations of the \textit{Virasoro algebra}, which is related to a central
extension of the Lie algebra of infinitesimal conformal transformations -
hence the term ``conformal'' field theory.

A $2D$\ CCFT with chiral algebra $\cala$\ can be reconstructed from the unitary
representations of $\cala$\ \cite{frki}. We need to recall the
notions of conformal weight and fusion rules which come from representation
theory.

Let $\la$\ be a unitary representation of $\cala$, and let 
$\calh_{\la}$\ denote the corresponding representation space, which is a
Hilbert space.
To each such representation one assigns a nonnegative number,
$\DE_{\la}$, called the \textit{conformal weight} of the representation. It is 
defined as the minimal eigenvalue of the zero-mode operator, $L_0$, of the
energy-momentum tensor, 
\begin{equation}
  \DE_{\la} = \inf
     \left\{
       \langle v_{\la},L_0v_{\la}\rangle
       \;|\;v_{\la}\in\calh_{\la},\|v_{\la}\|=1
     \right\} \, . 
\end{equation}
In a consistent theory there is always an irreducible \textit{vacuum
  representation}, $\om$, characterized by the vanishing of the conformal
weight, $\DE_{\om}=0$. 

Given two representations, $\la$\ and $\mu$, one can define their
{\em fusion}, namely a \textit{tensor
product representation}, $\la\ast\mu$, which is again a unitary
representation of $\cala$.
A chiral algebra is called \textit{rational} iff the number of inequivalent,
irreducible unitary representations is finite. Let us denote by $\LA$\ the set 
of such representations. For a rational chiral algebra, the tensor product of
two representations can be decomposed into a direct sum of irreducible unitary 
representations. 
Thus, the set of unitary irreducible representations
of a rational chiral algebra, furnished with the tensor product, has
the structure of a commutative, associative ring. 
For $\la_1,\la_2$\ and $\la_3$\ in $\LA$, let $N_{\la_1,\la_2}^{\la_3}$\
denote the multiplicity of $\la_3$\ as a subrepresentation in the tensor product 
$\la_1\ast\la_2$. The multiplicities $N_{\la_1,\la_2}^{\la_3}$\ are the
structure constants of the ring and are called
\textit{fusion rules}; for a rational chiral algebra, they are finite
non-negative integers. 
The vacuum representation, $\om$, plays the r\^ole of the unit for the
tensor product, i.e., $\la\ast\om=\om\ast\la=\la$. To every irreducible
representation
$\la$\ there corresponds a contragradient (or conjugate) representation
$\bar{\la}$\ with the property that $\la\ast\bar{\la}$\ contains the vacuum
    representation $\om$\ exactly once as a subrepresentation.

Given a number $n$\ of irreducible unitary representations,
$\la_1,\ldots,\la_n$, we define the linear space of \textit{conformal blocks}
as the space of invariant tensors, i.e., of invariant linear functionals, on
the representation space of the tensor-product representation
$\la_1\ast\cdots\ast\la_n$. It actually turns out (see \cite{fefk3})
that the tensor product
representation $\la_1\ast\cdots\ast\la_n$\ depends on $n$\ complex parameters
$z_1,\ldots,z_n$, which can be considered as coordinates of pairwise different 
points of the complex plane to which the cylinder can be mapped.
For this reason, the space
\begin{equation}
  V_{S^2}(z_1,\la_1,\ldots,z_n,\la_n)
\end{equation}
of conformal blocks depends on the complex parameters
$z_1,\ldots,z_n$. Its dimension is given by
\begin{equation}
  \caln_{\la_1,\ldots,\la_n} \;=\;
    \sum_{\mu_1,\ldots,\mu_{n-3}}
    N_{\la_1\la_2}^{\mu_1} N_{\mu_2\la_3}^{\mu_2} \ldots 
    N_{\mu_{n-3}\la_{n-1}}^{\bar{\la}_n}
  \quad,
\end{equation}
and does not depend on the parameters $z_1,\ldots,z_n$.

\begin{figure}[tb]
\begin{center}
\input{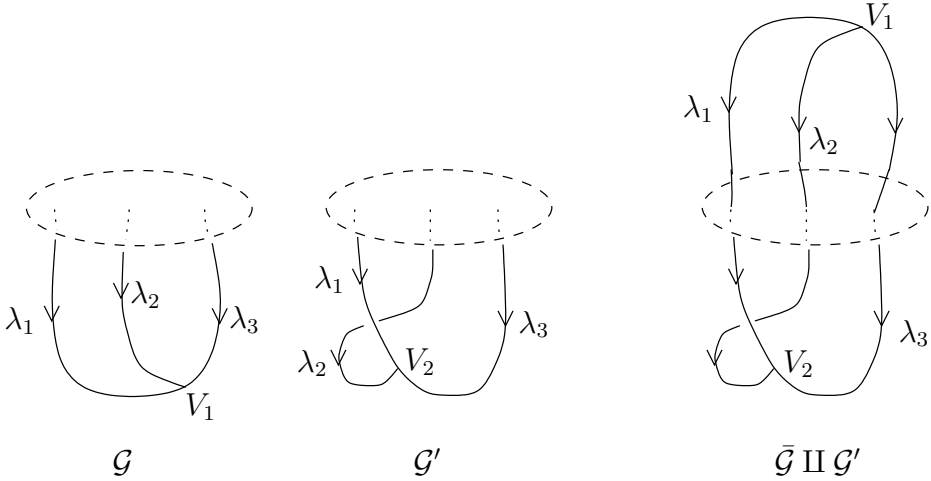}
  \parbox{14cm}{\caption{The ribbon graph $\bar{\calg}\amalg\calg'$\ obtained
  \label{fig:ribbonscalprod}
      from the ribbon graphs of $\calg$\ and $\calg'$.}}
\end{center}
\end{figure}

Next, we explain in which way 2d CCFT's arise in the description of
incompressible QHF's in the scaling limit.
We wish to describe the physical state space describing the scaling limit
of an incompressible QHF confined to a disc $D$, which for our purposes can be
viewed as a punctured two-dimensional sphere $S^2$, the boundary being mapped 
to $z=\infty$. It turns out that this state space can be identified 
with the space of conformal blocks of some CCFT! 
Let $\cala$\ be the chiral algebra characterizing a CCFT. The representations
of $\cala$\ are used as the quantum numbers labelling the static sources in
the bulk of the QHF. 
The boundary conditions can be described by vectors in a
representation space of the chiral algebra. 
Fixing a boundary condition
$v_{\la}\in\calh_{\la}$, and inserting static sources labelled by quantum
numbers $\la_1,\ldots,\la_n$\ corresponding to representations of $\cala$\ at
points $z_1,\ldots,z_n$\ in the disc $D$, the \textit{space of physical
  states} of the QHF is identified with the space of conformal blocks \cite{emss}
\begin{equation}
  V_{S^2}(z=\infty,\bar{\la},z_1,\la_1,\ldots,z_n,\la_n)[v_{\la}]
\end{equation}
with the vector $v_{\la}$\ inserted in the first argument (corresponding to
the point $z=\infty$). We denote this space by
$\calh_{\vec{z},\vec{\la}}[v_{\la}]$.
In order to select a specific vector in $\calh_{\vec{z},\vec{\la}}[v_{\la}]$\
and, in particular, to fix the generalized Aharonov-Bohm phases, we consider
an adiabatic evolution of sources in the QHF described by a ribbon graph,
$\calg$, with $n+1$\ external lines decorated by the representations
$\la,\la_1,\ldots,\la_n$\ ending at the points $\infty,z_1,\ldots,z_n$\
respectively; see Figure \ref{fig:ribbon2}.
\begin{figure}[tb]
\begin{center}
\input{ribb2.pstex_t}
  \parbox{14cm}{\caption{A ribbon graph for a state in
  \label{fig:ribbon2}
      $\calh_{\vec{z},\vec{\la}}[v_{\la}]$.}} 
\end{center}
\end{figure}
To each vertex of the ribbon graph, one associates a coupling, which
is an element of a linear space whose dimension is given by the corresponding
fusion rule (e.g., at the vertex $V_1$\ of Figure \ref{fig:ribbon2}, the
dimension is given by $N_{\la\bar{\la_2}}^{\mu}$). It is well known that
these data precisely specify a conformal block
\begin{equation}
  |\psi_{\calg}\rangle\;\in\;
  \calh_{\vec{z},\vec{\la}}(v_{\la})  
  \quad.
\end{equation}
It remains to describe the scalar product,
\begin{equation}
  \label{eq:tftamplitude}
  \langle\psi_{\calg}|\psi_{\calg'}\rangle
  \quad,
\end{equation}
of two vectors
$|\psi_{\calg}\rangle\;\in\;\calh_{\vec{z},\vec{\la}}(v_{\la})$\ and
$|\psi_{\calg'}\rangle\;\in\;\calh_{\vec{z},\vec{\la}}(w_{\la})$. This scalar
product is given by
\begin{equation}
  I_{\bar{\calg}\amalg\calg'}\cdot\langle v_{\la},w_{\la} \rangle
  \quad,
\label{eq:neu}
\end{equation}
where $\langle\cdot,\cdot\rangle$\ denotes the scalar product in the
representation space $\calh_{\la}$, and $I_{\bar{\calg}\amalg\calg'}$\ is the
invariant the three-dimensional topological field theory assigns to the ribbon 
graph obtained by gluing the reflected version $\bar{\calg}$\ of the ribbon graph
$\calg$\ to the ribbon graph $\calg'$\ at the end points of the external
lines, and $\bar{\calg}$\ is the ribbon graph obtained from $\calg$\ by
reversing the orientation of the lines of $\calg$; see Figure
\ref{fig:ribbonscalprod}. 
The invariant $I_{\bar{\calg}\amalg\calg'}$\ is a generalized Aharonov-Bohm
phase and can be calculated from the data
of the underlying 2d CCFT (the fusion rules, the fusing and braiding
matrices,\ldots); see \cite{frki}.
The three-dimensional theory with transition amplitudes given by
\erf{eq:tftamplitude}, \erf{eq:neu} is called a \textit{topological} field 
theory, because
these transition amplitudes \textit{only} depend on the \textit{topology} of
$\bar{\calg}\amalg\calg'$, and not on the precise way in which
$\bar{\calg}\amalg\calg'$\ is embedded into three-dimensional space-time.

The explicit expressions for the vectors $|\psi_{\calg}\rangle$, in particular 
their dependence on the insertion points $z_1,\ldots,z_n$, may remind one of
generalized Laughlin ansatz wave functions \cite{laufra,gipr}. 
However, this resemblance is largely accidental (and gauge-dependent)! 
Conformal blocks, both of unitary and non-unitary \cfts, can provide
a useful description of `special functions'  and have been used for this
purpose also in other situations, e.g.\ for the prepotential in Seiberg-Witten
models \cite{floh8} and for the description of wave functions for the BCS
Hamiltonian \cite{sieR2}.
No connection of this kind will be invoked in this paper!

\section{Conditions on a chiral conformal field theory describing a Quantum
  Hall Fluid}
  \label{sec:conditions}

Thanks to the correspondence between a class of $3D$\ TFT's and $2D$\ CCFT's
discussed in  Section \ref{sec:generalfeatures}, we can study theories
describing incompressible QHF from the point of view of CCFT, instead of TFT. 
In the following, we describe properties that characterize a CCFT whose
corresponding TFT can be expected to be relevant in
describing an incompressible QHF; we will denote these CCFT's as quantum
Hall CCFT's. 

\subsection{Consistency conditions}\label{3.1}

To define a CCFT, we must specify a chiral algebra $\cala$\ and a set $\LA$\
of \textit{unitary} irreducible representations containing a unique vacuum
representation $\om$\ (which is characterized by the vanishing of the
conformal weight $\DE_{\om}=0$). 

As observed in Section \ref{sec:generalfeatures}, the chiral algebra of a
Quantum Hall CCFT must contain a $\uone$-current algebra.
This means that the static pointlike excitations of the bulk carry an additive 
quantum number, which is interpreted as the electric charge of the source.
For simplicity, and because we are not attempting a complete classification, we
assume that the chiral algebra is a direct product
\begin{equation}
  \label{eq:chiralalgebra}
  \cala=\calc\otimes\uone
  \quad,
\end{equation}
where $\calc$\ is an \textit{electrically neutral} chiral algebra. Let $\PI$\
denote the set of unitary irreducible representations of $\calc$, which is
closed under fusion. We furthermore assume that $\calc$\ is \textit{rational},
i.e., that $\PI$\ is finite. Physically, this means that, for a fixed electric
charge, there are only finitely many different static pointlike sources, also
called quasi-particles, that carry that charge.

Before going on, let us recall some facts concerning the $\uone$-theory that
are needed later. 
The unitary irreducible representations are labelled by
``charges'', i.e., by real numbers $r$; the conformal weights are given by 
$\DE_r=\frac{r^2}{2}$, and the corresponding fields are 
vertex operators, $:e^{ir\phi}:$, which are Wick-ordered exponentials of a
massless, chiral free field $\phi$\ . 
The electric current $j$\ is expressed in terms of $\phi$\ by 
$j=\sqrt{\si_H}\partial\phi$; it follows that the electric charge,
$\qq$, of a representation is $\qq_r=\sqrt{\si_H}r$.

The following consistency conditions, (C1) through (C5), for the pair
$(\cala,\LA)$\ reflect physical principles and pragmatic considerations that
have proven to be successful in a previous classification of incompressible
QHF based on abelian current algebra; see \cite{frst}.
\begin{itemize}

\item[(C1)] \textit{Physical representations}\\
Unitary representations of $\cala$\ are constructed as tensor products of
representations of $\calc$\ and of $\uone$. Let us denote by $\LA$\ the set of
physically realized representations. Since $\cala$\ is a direct product, we have
\begin{equation}
  \label{eq:repinclusion}
  \LA \quad \subseteq \quad \PI \times \reals \quad,
\end{equation}
which means that representations of $\cala$\ are of the form $l=(\pi,r)$,
with $\pi\in\PI$\ and $r$\ a real number.
If two pointlike excitations meet at the same point in the bulk, they
generate another pointlike excitation of the fluid.
This is a way to express the requirement that the set, $\LA$, of
representations be closed under fusion.

\item[(C2)] \textit{Existence of one-electron states}\\
Among the physically realizable representations there should be (at
least) one representation $e$\ with electric charge $-1$. 
A pointlike source labelled by $e$\ represents an electron that has been
inserted somewhere in the bulk. We denote the corresponding representation by
$e=(\vep,r_e)$.
To say that $e$\ has electric charge $-1$\ means that
\begin{equation}
  \label{eq:eleccharge}
  \sqrt{\si_H}\;r_e=-1
  \quad.
\end{equation}

We thus require that there be a nonempty family of representations
$\LAE=\{e_a|a=1, \ldots, \nel\}$\ in $\LA$\ satisfying \erf{eq:eleccharge} 
which we call (one-) electron representations.

We then define a family $\LAM$\ in $\LA$\ of multi-electron representations as 
the representations obtained by multiple fusing of representations in
$\LAE$. This means that a pointlike source labelled by such a representation is 
obtained by letting several electron sources coalesce in one single point,
generating a multi-electron cluster.
The electric
charge of representations in $\LAM$\ can be determined by making use of the
fact that the electric charge is an {\em additive} quantum number under fusion.

\item[(C3)] \textit{Charge and statistics}\\
Consider the state corresponding to a single pointlike source $\la$\ in the bulk.
If the sample is rotated by $2\pi$\ with respect to an axis perpendicular to
the sample plane, then the resulting vector differs from the initial one by a
phase $e^{2\pi i \DE_{\la}}$, where $\DE_{\la}$\ is the conformal weight of the
representation $\la$. 
If $\DE_{\la}$\ is an integer, then $\la$\
is said to obey Bose-statistics; if $\DE_{\la}$\ is half-integer, then $\la$\ is said to obey Fermi-statistics; if $\DE_{\la}\neq0\mbox{ mod
  }\frac{1}{2}$, then $\la$\ obeys fractional statistics.

We require that those excitations of the incompressible QHF that have been identified with
multi-electron pointlike sources obey Fermi/Bose-statistics depending on whether
they contain an odd or an even number of electrons, respectively; i.e., we 
require that, for the multi-electron representations $m\in\LAM$, the following
\textit{charge-statistics connection} holds:
\begin{eqnarray}
  \label{eq:chargestatistic}
  \qq_m=0\mod 2 & \quad \Longrightarrow \quad &
     \DE_m=0\mod 1\qquad \mathrm{(Bose}\;\mathrm{
       statistic)} \nonumber \\
  \qq_m=1\mod 2 & \quad \Longrightarrow \quad &
     \DE_m=\frac{1}{2}\mod 1\qquad\mathrm{(Fermi}\;\mathrm{
       statistic)}
     \quad.
\end{eqnarray}
Here, $\qq_m$\ denotes the electric charge of the multi-electron 
representation $m$.

\item[(C4)] \textit{Relative locality}\\
States with a multi-electron cluster in the bulk should be single valued
functions of the position of that cluster. This requirement, which goes under
the name of \textit{relative locality}, means that when a
multi-electron pointlike source is moved along a closed path in the bulk,
possibly winding around other static pointlike sources, then the final state 
vector is the same as the initial state vector.
It turns out that this requirement can be expressed in terms of fusion rules 
and conformal weights as follows:
\begin{equation} \label{relloc}
 \mbox{For all } \la,\la'\in\LA,\;m\in\LAM, 
 \mathrm{with}\ N_{\la m}^{\la'}\neq0\,,
 \quad
 \DE_{m}+\DE_{\la}-\DE_{\la'}=0\mod 1
 \quad.
\end{equation}

\item[(C5)] \textit{Charge and spin}\\
If spin is a nontrivial quantum number, then, in multi-electron states, it
should be determined by the spins of the electrons. Spin labels the
representations of an $\sutwo_k$\ current algebra in the electrically
neutral factor $\calc$\ of $\cala$. If we identify a subalgebra
$\sutwo_k\in\calc$\ as describing spin, then we require the multi-electron
representations  $m\in\LAM$\ to obey a \textit{spin and charge connection}:
\begin{equation}
  \label{eq:chargespin}
\begin{split}
  \qq_m=0\mod 2 & \Longrightarrow 
                s_m=0\mod 1 \\
  \qq_m=1\mod 2 & \Longrightarrow 
                s_m=\frac{1}{2}\mod 1 \\
\end{split}
\end{equation}
Here, $s_m$\ denotes the $\sutwo$-spin of the representation $m$.

\end{itemize}

At this point, an important remark should be made: From the assumption
that $\calc$ be rational and from condition (C3) 
it follows that the Hall conductivity is a
\textit{rational number}. This can be seen as follows: The conformal weight of
an electron representation is given by
\begin{equation}
  \label{eq:elecweight}
  \DE_e\;=\;\frac{r_e^2}{2}+\DE_{\vep}\;=\;\frac{1}{2\si_H}+\DE_{\vep}
\end{equation}
and, by (C3), it must be half-integer, say $\frac{1}{2}+j$\ with $j$\ a
positive  integer. 
It is known that, for a rational CCFT, the conformal weights are
rational numbers. Thus, $\DE_{\vep}$\ in \erf{eq:elecweight} is a rational
number. It then follows that the Hall conductivity 
\begin{equation}
  \label{eq:hallrational}
  \si_H\;=\;\frac{1}{1+2j- \DE_{\vep}}
\end{equation}
is \textit{rational}.

Next, we define a simple, but useful transformation, called the 
\textit{shift map} (compare \cite{frst}), that maps a Quantum Hall CCFT to 
another Quantum Hall CCFT modifying only the electrically charged part
of the theory: the electric current is transformed as 
\begin{equation}
  j'\;=\;\sqrt{\frac{\si_H}{1+2p\si_H}}\;j
  \quad,
\end{equation}
where the shift parameter $p$\ is a positive integer. The Hall
conductivity is transformed as 
\begin{equation}
  \si_H'\;=\;\frac{1}{1+2p\si_H}\;\si_H
  \quad.
\label{eq:34}
\end{equation}
The $\uone$-label of the representations is scaled in such a way that the 
electron representations continue to have charge equal to 1: 
$r_e'=r_e\sqrt{1+2p\si_H}$. That is, the shift map relates an incompressible
QHF with Hall conductivity $\sigma_H$ to a {\em putative} QHF with
Hall conductivity $\sigma_H'$ given by \erf{eq:34} by mapping the fusion ring
$\Lambda$ of the chiral algebra $\cala=\calc\otimes \hat u(1)$ to a fusion ring
$\Lambda'$ of the same chiral algebra as follows:
\begin{equation}
  \Lambda \ni \la=(\pi,r) \quad \longmapsto \quad \la'=(\pi,r'=r\sqrt{1+2p\si_H})
\in \Lambda' \, . 
\label{eq:35} \end{equation}
Usually, $\Lambda'$ arises from the image of $\Lambda$ under the map 
introduced in \erf{eq:35} by adding further fractionally charged representations
(keeping the set of representations of $\calc$ fixed). 
The only restriction comes from 
the requirement that the multi-electron fields are relatively local with
respect to the fields corresponding to the new representations.

\subsection{Remarks on modular invariance and covariance}\label{3.2}
We wish to comment on the r\^ole that the modular group $SL(2,\zet)$ may
play in the analysis of incompressible QHF's.

\begin{figure}[!htb]
\begin{center}
\input{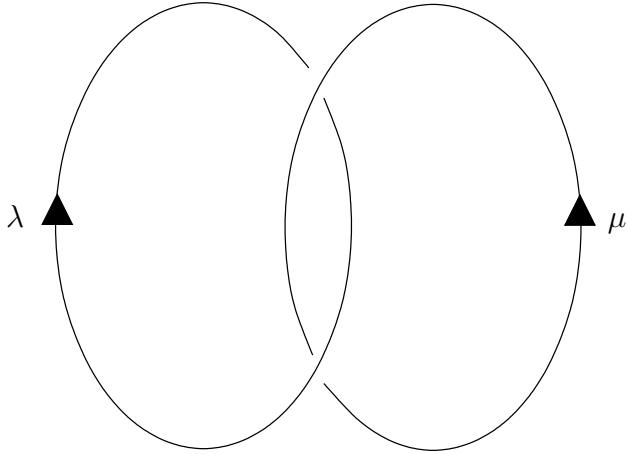}
  \parbox{14cm}{\caption{The link whose link invariant in the three sphere
       $S^3$ equals   \label{fig:s}
      $S_{\lambda\mu}$.}}
\end{center}
\end{figure}

For any TFT, one can consider the link in Figure \ref{fig:s}
in the three-sphere for any pair of representations $\lambda$ and $\mu$
and compute the corresponding link invariant $S_{\lambda\mu}$, which is
a complex number. Topological invariance implies that the matrix $S$
with matrix elements $S_{\lambda\mu}$ is symmetric. Experience shows that
requiring this matrix to be invertible ensures {\em completeness} of the
theory, i.e., it ensures that one has included all types of static
pointlike sources.

Let us assume that the fractional part of the conformal weights in each
superselection sector for $\cala$ are constant. This assumption allows
us to define a diagonal unitary matrix $T$ by
\begin{equation}
T_{\lambda\mu} = \delta_{\lambda\mu} \exp(2\pi\ii(\Delta_\mu-c/24)) \, . 
\end{equation}

Surgery operations for links in three-manifolds can be used to show that the 
matrices $S$ and $T$ generate a unitary representation of the modular
group. If the fractional part of the conformal weight $\Delta$ is not constant, 
but the fractional part of $N\Delta$ is constant for all states in
a given superselection sector, one still finds a representation of the
subgroup of the modular group generated by $S$ and $T^N$.
It turns out that, in the description of incompressible QHF's in terms of
CCFT's, it is quite natural to require covariance of the fusion ring under the
subgroup of the modular group corresponding to $N=2$; see Appendix 
\ref{app:modular}.

We emphasize that the requirement of covariance of the fusion ring under 
subgroups 
of the modular group is formulated completely on the level of {\em chiral}
CFT. It should not be confused with modular {\em in}variance of a torus
partition function. The latter is a requirement in {\em full} CFT, which
is a theory in which left movers and right movers have been combined.
For the description of QHF's, chiral CFT is relevant; the consideration
of partition functions that are invariant under subgroups of the modular group
does not occur naturally.

\subsection{Stability of the Incompressible Quantum Hall Fluid}\label{3.3}
Experimentally, only those incompressible QHF are accessible which are stable under small
changes of the experimental control parameters, like, for example, the shape
of the sample, the concentration of impurities or small inhomogeneities in the
external magnetic field. 
This raises the question of how to assess the
stability of an incompressible QHF described by a given TFT (and corresponding CCFT). In
this paper, we shall not present an answer to this question based on an
analysis of the microscopic quantum theory of incompressible QHF's. 
Instead, we propose some stability criteria extracted from the comparison of
experimental data with theoretical predictions made in the framework of Quantum 
Hall Lattices \cite{frst}.
These data indicate that an incompressible QHF is the more stable $\ldots$
\begin{itemize}
\item[(S1)] $\dots$\ the smaller the central charge $c_{\cala}=1+c_{\calc}$\ is.
\item[(S2)] $\dots$\ the smaller the conformal weights $\DE_e$\ of the
  electrons are.
\item[(S3)] $\dots$\ the smaller the number $\nfra$\ of
  representations with electric charge $0\leq\qq<1$\ is.
\end{itemize}
Concerning (S2) we remark that, experimentally, no incompressible QHF wit
$\si_H<\frac{1}{7}$\ has been observed. The conformal weights of the electrons 
are bounded from below by $\DE_e\geq\frac{1}{2\si_H}$, a consequence of
\erf{eq:elecweight}. This motivates the theoretical speculation that, for
$\si_H<\frac{1}{7}$, or $\DE_e>\frac{7}{2}$, the ground state of the
system is a Wigner crystal, which is obviously not an incompressible state
because of the existence of gapless modes (phonons). The criterion (S2) could
then be interpreted as follows: the incompressible QHF is the more stable, the
``farther remote'' its groundstate is from such a crystalline ground state.

\section{Quantum Hall Lattices}
\label{sec:lattices}
In this section we review a description of incompressible QHF's in terms of
a class of CCFT's whose chiral
algebra is (an extension of) a $\uone^N$-current algebra. 
This situation has been studied extensively, and has led to a partial
classification of incompressible QHF in terms of Quantum Hall Lattices (QHL);
see \cite{frst}.

Representations of a $\uone^N$-current algebra are labelled by points $\rr$\
in euclidean $\reals^N$. The conformal weights are
$\DE_{\rr}=\frac{\rr^2}{2}$, and the corresponding primary fields are
Wick-ordered exponentials of the form
\begin{equation}
  \PSI_{\rr}\;=\;:e^{i\sum_{j=0}^N r^j\phi_j}:
  \quad.
\end{equation}
The real numbers $(r^i)_{i=1..N}$\ are the components of $\rr$\ with respect 
to an
orthonormal basis; the $\phi_i$\ are massless chiral free fields with
$\partial\phi_i=j_i$, where the $j_i$\ are the currents that generate the
$\uone^N$-current algebra with anomalous commutators
\begin{equation}
  \label{eq:u1Ncurrentalg}
  [j_{i,m},j_{j,n}]\;=\;\de_{m+n,0}\;\de_{ij}
  \quad.
\end{equation}
The fusion product of representations is simply $\rr\ast\rr'=\rr+\rr'$.

In \cite{frst}, incompressible QHF's are described in terms of
$\uone^N$-current algebras. It is shown that:
\begin{itemize}
\item[(a)] There is an odd integral lattice $\GA$\ in euclidean $\reals^N$,
  with scalar product denoted by $\langle.,.\rangle$, such that the set of
  physically realized representations is a lattice $\GA_{\mathrm{phys}}$\ with
\begin{equation}
  \GA\subseteq\GA_{\mathrm{phys}}\subseteq\GA^{\ast}\quad.
\end{equation}
Here $\GA^{\ast}$\ denotes the dual lattice of $\GA$.
\item[(b)] The electric current is $j=\sum_iQ^ij_i$, where $(Q^i)_{i=1..N}$\ 
     are shown to be the components with respect to the chosen orthonormal 
    basis of an \textit{odd,
    primitive} vector $\QQ\in\GA^{\ast}$. 
 ``Odd'' means that for any vector $\rr\in\GA$\ we have
  $\langle\QQ,\rr\rangle=\langle\rr,\rr\rangle\mod2$; 
  ``primitive'' means that if it is joined to the origin by a line segment the
   latter does not contain any other point in $\GA^{\ast}$.
\end{itemize}

A CCFT with a $\uone^N$-current algebra as chiral algebra and describing an
incompressible QHF is therefore determined by a pair $(\GA,\QQ)$ with the
properties described in (a) and (b). The Hall conductivity is given by
\begin{equation}
  \si_H=\langle\QQ,\QQ\rangle
  \quad.
\end{equation}
In the following we show that these data are equivalent to a
Quantum Hall CCFT, $(\cala,\LA)$, that fulfills the consistency
conditions of Section \ref{sec:conditions}.
\begin{itemize}
\item[(C1)] The set $\{j=\sum_iK^ij_i\in\uone^N| \langle \QQ,\KK\rangle=0\}$\ 
  generates an electrically neutral $\uone^{N-1}$-current algebra. 
  To obtain a rational theory in 
  the electrically neutral sector, we pass to a simple current extension of
  $\uone^{N-1}$\ with uncharged fields of integer conformal weight. 
  This amounts to defining $\calc$ as the chiral algebra generated by
\begin{equation}
  \label{eq:neutralalglatt}
  \left\{j=\sum_iK^ij_i\in\uone^N| \langle \QQ,\KK\rangle=0\right\}
  \;\cup\;
  \left\{:e^{i\sum_ir^i\phi_i}:|\rr\in Q^{\perp}\right\}
  \quad,
\end{equation}
where
\begin{equation}
  Q^{\perp}\;=\;
  \left\{\rr\in\GA|
      \langle\QQ,\rr\rangle=0 \right\} \quad.
\end{equation}
Note that, by property (b), $\langle \rr,\rr\rangle$ is {\em even}, for every
$\rr\in Q^\perp$.
The unitary representations of such $\calc$\ are labelled by points in
\begin{equation}
  \PI\;=\;
      (Q^{\perp})^{\ast}\big/Q^{\perp}
  \quad,
\end{equation}
with
\begin{equation}
  (Q^{\perp})^{\ast}\;=\;
         \left\{\rr^{\perp}|
          \langle\QQ,\rr^{\perp}\rangle=0;
          \langle\nn,\rr^{\perp}\rangle=0\mod1
          \;\forall\nn\in Q^{\perp}\right\}
  \quad.
\end{equation}

\item[(C2)] The set of electron representations can be identified with
\begin{equation}
  \LAE\;=\;
  \left\{\rr_e\in\GA|\langle\QQ,\rr_e\rangle=1\right\}\big/Q^{\perp}
  \,.
\end{equation}
Note that the number of points in $\LAE$ is usually larger than 1. This means
that, in general, there are several species of electrons distinguished
from each other by some ``internal quantum numbers''.
The set of multi-electron representations is then given by 
\begin{equation}
  \LAM\;=\;
  \left\{\rr_m\in\GA|\langle\QQ,\rr_m\rangle=j,j\in\naturals,j\geq1\right\}
         \big/Q^{\perp} 
  \quad.
\end{equation}
Multi-electron representation spaces are direct sums of countably many
representation spaces of the $\uone^N$-current algebra:
\begin{equation}
  \calh_m=\bigoplus_{\ss\in Q^{\perp}}\calh_{\rr_m+\ss}
  \quad.
\end{equation}
The electric charge of such representations is a 
positive integer, $\qq_m=\langle\QQ,\rr_m\rangle$, and is $1$\ for electrons.
The set $\LAM$\ can be obtained by multiple fusion of representations in
$\LAE$.

\item[(C3)]
 The conformal weights of multi-electron representations are
\begin{equation}
  \DE_m=\mathrm{min}_{\ss\in Q^{\perp}}
                     \left\{\DE_{\rr_m+\ss}\right\}
  \quad.
\end{equation}
The charge and statistics connection is fulfilled, as follows from the
definition of oddness of $\QQ$:
\begin{equation}
  2\DE_m\;\equiv\;\langle\rr_m,\rr_m\rangle\;=\;\langle\QQ,\rr_m\rangle\;=\;\qq_m\mod2
  \quad.
\end{equation}

\item[(C4)] 
The complete set of representations is
\begin{equation}
  \LA\;=\;\GA_{\mathrm{phys}}\big/Q^{\perp}
  \quad,
\end{equation}
and each representation is a direct sum of countably many
representations of the $\uone^N$-current algebra:
\begin{equation}
  \calh_{\la}=\bigoplus_{\ss\in Q^{\perp}}\calh_{\rr_{\la}+\ss}
  \quad.
\end{equation}
Again, the electric charge of such representations is given by
$\qq_{\la}=\langle\QQ,\rr_{\la}\rangle$.
The fusion product of such representations corresponds to addition mod
$Q^{\perp}$. The relative locality condition for the multi-electron fields reads
\begin{equation}
  \DE_e+\DE_{\la}-\DE_{e\ast\la}=-\langle\rr_e,\rr_{\la}\rangle=0\mod1 \,,
\end{equation}
for all $e\in\LAE$\ and $\la\in\LA$. It is fulfilled, since $\rr_e\in\GA$\ and 
$\rr_{\la}\in\GA^{\ast}$.

\item[(C5)] The electrically neutral algebra $\calc$ defined through
  \erf{eq:neutralalglatt} contains an $\sutwo$-current algebra at level 1
  iff the lattice $Q^{\perp}$\ factorizes into sublattices
  and one of these factors is the root lattice of $\sutw$. This can occur
  in the case of ``maximally symmetric QHL's'':
   A maximally symmetric QHL is denoted
by $(L|^{\al}\gg)$: $\gg$\ is a semisimple Lie algebra, $\al$\ a minimal
weight thereof and $L$\ is an odd positive integer, with $L>(\al,\al)$; 
the QHL  is generated by the root lattice of $\gg$\ and by
$\rr_e=\al+(L-(\al,\al))\QQ$, where $\QQ$\ is the charge vector perpendicular
to the root lattice of $\gg$.
There is a unique electron representation, $e$, which corresponds to
$\rr_e$. If $\gg$\ has an $\sutw$-factor then the electron must have
$\sutwo$-spin $\frac{1}{2}$, since $\al$\ is minimal. 
The charge and spin
connection in such a case is seen to hold for the electron representation; 
for  multi-electron representations, that connection follows from the
additivity of the electric charge under fusion and from the additivity mod
$2$\ of $\sutwo$-spin. 

\end{itemize}

The parameters that enter in the stability criteria are connected to lattice
invariants:
\begin{itemize}
\item[(S1)] The central charge of the Quantum Hall CCFT is the rank, N, of the 
    lattice.
\item[(S2)] The largest conformal weight of the electrons is bounded from
  above by a lattice invariant called $l_{\mathrm{max}}$; see \cite{frst} for
  more detailed explanations.
\item[(S3)] The number of representations with fractional electric charge
  is bounded from above by the discriminant, $|\GA^{\ast}/\GA|$, of
  the lattice.
\end{itemize}

\section{Construction of chiral conformal theories describing incompressible
Quantum Hall fluids}
\label{sec:recipe}

The goal of this section is to describe a simple algorithm for the construction
of Quantum Hall CCFT's.
\subsection{Explicit construction with electrons as simple currents}
The main assumption underlying our algorithm is that electrons are 
\textit{simple currents}; for a review on simple currents see \cite{scya6}. 

An irreducible
representation of  a chiral algebra is called a simple current if its fusion
with any other irreducible representation yields exactly one irreducible
representation. Equivalently, a simple current $e$\ is characterized by the
fusion rule
\begin{equation}
 e\ast\bar{e}=\om \,,
 \quad.
\end{equation}
where $\bar e$ is the representation conjugate to $e$.
Thus our assumption simply means that when a hole is filled with
a corresponding electron, the state of the system is the vacuum state. 
When dealing with simple currents, a useful concept is that of
\textit{monodromy charge}. The monodromy charge (or simply the monodromy) of a
representation $\la$\ with respect to a simple current $e$\ is given by
\begin{equation}
 Q_e(\la)=\DE_e+\DE_{\la}-\DE_{e\ast\la} \mod1
 \quad.
\end{equation}
If $e,e_1,e_2$ are simple currents, $\la,\la_1,\la_2$\ arbitrary representations,
we have that 
\begin{eqnarray}
 Q_{e_1}(e_2) & = & Q_{e_2}(e_1) \label{eq:monodromy1}\\
 Q_e(\la_1\ast\la_2) & = & Q_e(\la_1)+Q_e(\la_2) \mod1 \label{eq:monodromy2}\\
 Q_{e_1\ast e_2}(\la) & = & Q_{e_1}(\la)+Q_{e_2}(\la)
        \mod1 \label{eq:monodromy3}
 \quad.
\end{eqnarray}
The set of simple currents of a theory, endowed with the fusion product $\ast$, 
has the structure of an abelian group, with unit corresponding to the vacuum 
representation $\om$. We note that a simple current is relatively local iff 
all representations occurring in the theory have vanishing monodromy
with respect to the simple current.

The assumption that electron fields are described by simple currents
drastically simplifies the analysis of the consistency conditions (C3) and 
(C4) of Section \ref{3.1}. In the following, we show how to construct a 
Quantum Hall CCFT satisfying conditions (C1) through (C5) and (S1) through
(S3) (see Section \ref{3.1} and \ref{3.3}) with electron fields given by simple
currents. 
\begin{itemize}
\item[(i)] The electron representations, $\LAE$, are of the form
  $e_a=(\vep_a,\frac{1}{\sqrt{\si_H}})$, where $\vep_a$\ are simple currents 
  of the $\calc$-theory, with the property that \\
  $\bullet$\ the charge-statistics connection (C1) is satisfied for each
  electron, i.e., 
\begin{equation}
  \label{eq:recipeweight}
  \frac{1}{\si_H}+2\DE_{\vep_a}\;=\;1 \mod 2, \qquad a=1, \ldots, \nel
\end{equation}
where $\nel$ is the number of distinct species of electrons; and \\
  $\bullet$\ the relative locality condition is satisfied for each pair of
  electrons
\footnote{If there is only one electron, then \erf{eq:recipemonodromy1}
  follows directly from \erf{eq:recipeweight}. This can be proven using the
  identities $2\DE_{\vep}=\frac{r(N-1)}{N}$\ and $Q_{\vep}(\vep)=\frac{r}{N}$, 
  where $N$\ is the order of $\vep$, and $r$\ is an integer defined mod $N$.},
  i.e.,
\begin{equation}
  \label{eq:recipemonodromy1}
  Q_{e_a}(e_b)\;=\;-\frac{1}{\si_H}+Q_{\vep_a}(\vep_b)\;=\;0\mod 1, 
         \qquad a,b=1,\ldots,\nel
\end{equation}
\item[(ii)] As a consequence of (i), the multi-electron representations, 
  $\LAM$, obtained by multiple
  fusion of electron representations are again simple currents. The relative
  locality condition (C4), $Q_{m_1}(m_2)=0$, for $m_1,m_2\in\LAM$, is
  fulfilled, as follows from \erf{eq:monodromy1}, \erf{eq:monodromy2} and
  \erf{eq:monodromy3}. 
  By induction, it is possible to prove that the charge-statistics
  connection (C3) is fulfilled, because from 
  $\DE_{m_1}=\frac{1}{2}\qq_{m_1}\mod1$,\ and $Q_{m_1}(m_2)=0$, it
    follows that $\DE_{m_1\ast m_2}=\frac{1}{2}\qq_{m_1\ast
      m_2}\mod1$.
\item[(iii)] Further representations, $\la=(\pi,r)$, have to fulfill the
  relative locality condition (C4). This amounts to requiring the vanishing of 
  the monodromy charges with respect to the electrons, i.e.,
\begin{equation}
  \label{eq:recipemonodromy2}
 Q_{e_a}(\la)\;=\;-\frac{\qq_r}{\si_H}+Q_{\vep_a}(\pi)\;=\;0 \bmod 1
\qquad a=1,\ldots,\nel
 \quad.
\end{equation}
  The vanishing of the monodromy charge with respect to multi-electron
  representations trivially follows from \erf{eq:monodromy3}.
  If $\LA$\ is defined to consist of \textit{all} representations $\la$\
  satisfying \erf{eq:recipemonodromy2} then $\LA$\ is closed under
  fusion. From \erf{eq:monodromy2} it indeed follows that fusion of two
  representations with vanishing monodromy charge with respect to a simple 
  current gives representations with vanishing monodromy charge with respect 
  to the (same) simple current.
\end{itemize}

Equation \erf{eq:recipemonodromy2} defines, for each representation $\pi$\ of
$\calc$, a discrete set of charges 
\begin{equation}
  \qq\;=\;\si_H Q_{\vep}(\pi)+\si_H k,\quad k\in \zet
  \quad,
\end{equation}
or, equivalently, a discrete set of (labels of) $\uone$-representations
\begin{equation}
  r\;=\;\sqrt{\si_H} Q_e(\pi)+\sqrt{\si_H} k,\quad k\in \zet
  \quad.
\end{equation}
Since $\si_H$\ is a rational number and $\calc$\ is rational, it is 
possible to find an even integer, $\caln$, such that all $\uone$-labels that
occur in $\LA$\ are of the form $r=\frac{l}{\sqrt{\caln}}$, for a suitable
integer $l$. 
The set of physically realizable representations is contained in
\begin{equation}
  \LA\;\subseteq\;\PI\times\zet\big[\frac{1}{\sqrt{\caln}}\big]
  \quad,
\end{equation}
as required by condition \erf{eq:recipemonodromy2}.

In general, a large amount of mathematical information can be obtained 
from the data of a CCFT. However, only a small part of that information
can be related to {\em experimentally accessible} physical quantities of an 
incompressible QHF. Among such physical quantities is the {\em minimal 
electric charge of a quasi-particle} of the QHF \cite{glat,rezn}. 
We can compute this quantity 
very easily as follows.

For simplicity, we assume that there is only one species of electrons present
in the theory. The neutral factor, $\vep$, of the electron field is a 
simple current of the $\calc$
theory. According to \cite{scya6}, the conformal weight of any simple current
$\vep$ is related to its order, $\ord{\vep}$, by an equation of the 
form:
\begin{equation}
2\Delta_{\vep} = \frac{\ord{\vep}-1}{\ord{\vep}} r_{\vep} \mod \zet
\end{equation}
where $r_{\vep}$ is an integer-valued quantity associated with
the simple current, defined
modulo $\ord{\vep}$. From equation \erf{eq:recipeweight}, writing
$\sigma_H=n_H/d_H$ for relatively prime integers $n_H$ and $d_H$, we conclude 
that $\ord{\vep}$ must be a multiple of $n_H$, 
\begin{equation}
\ord{\vep} = \ell n_H 
\label{64}
\end{equation} 
Here, $\ell$ must be a divisor of the quantity $r_{\vep}$. The
monodromy charge of any other representation, $\pi$, with respect to 
$\vep$ is of the form $r_{\pi}/\ord{\vep}$, \cite{scya6}, for some 
integer $r_{\pi}$. Combining this fact with \erf{64} and with 
\erf{eq:recipemonodromy2}, we conclude that the smallest possible electric 
charge is 
\begin{equation}
\qq_{\rm min}\; = \frac{1}{\ell d_H}
\label{eq:qmin}
\end{equation}
We cannot, of course, assert that this smallest possible charge can be
realized experimentally. An equation similar to \erf{eq:qmin} has been
derived for Quantum Hall lattices in \cite{frst}, where the analogue of $\ell$
was called {\em charge parameter}.

\subsection{Coset and orbifold construction of Quantum Hall CCFT's}
  \label{sec:coset}
Recently, the application of the {\em coset construction} to the quantum Hall 
effect has attracted some attention \cite{cagt}, 
because it provides a possibility to relate the
maximally symmetric QHL $(1|^{\al}\sutw\oplus\sutw)$, with $\si_H=\frac{1}{2}$,
$c_\cala=3$, to a $\vir_1\times\uone$-theory ($\vir_1$\ being the chiral 
algebra of the Ising model), with $\si_H=\frac{1}{2}$, $c_\cala=\frac{3}{2}$.

The relevance of the coset construction (see Appendix \ref{app:cosets}) for our
framework is twofold. First, the coset construction allows one to construct a
new class of theories, starting from WZW-models, always lowering the central
charge. In view of stability criterion (S1), the coset theories provide good
candidates for the theory corresponding to the electrically neutral chiral 
algebra $\calc$. Second, if 
there are gauge symmetries present in the theory corresponding to the 
incompressible QHF, which may be expected on physical grounds in a
particular situation, the coset construction is a method to implement the
gauge reduction: coset CCFT can be viewed as WZW-models in which a subgroup is 
gauged. 

One might want to also consider {\em orbifold theories} as candidates for 
the description
of incompressible QHF's. In view of the stability conditions of Section \ref{3.2}
they do not appear to be particularly promising candidates, though: in contrast
to the coset construction (see Appendix \ref{app:cosets}), the orbifold 
construction does {\em not} lower the Virasoro central charge. Moreover, the
orbifold construction requires additional fields, so-called twist fields,
for any primary field that is symmetric under the action of the orbifold group.
(For a discussion in the case when the orbifold group is $\zet_2$, see
e.g.\ \cite{bifs}).  Unless there are plenty of primary fields that are 
not symmetric under the action of the orbifold group, the orbifold theory
will therefore have more primary fields than the original theory. Thus,
in general, the orbifold theory can be expected to be less stable than the 
original theory.

\section{Examples of Quantum Hall fluids with 
         $\sigma_H=\frac{1}{2}\frac{e^2}{h},\frac{3}{5}\frac{e^2}{h},
         \frac{e^2}{h},\ldots$}
  \label{sec:examples}
  \renewcommand{\arraystretch}{1.5}

In this section, we illustrate the construction proposed in Section 
\ref{sec:recipe} by some simple but important examples.
Candidates for the electrically neutral theory $\calc$\ are the Virasoro
minimal models, simple current extensions thereof, and low-rank, low-level
WZW-models. (WZW models at level 1 are encountered in connection with
maximally symmetric QHL, as discussed  in Section \ref{sec:lattices}.)
These three classes of examples have small central charge, simple 
currents with small conformal weight defining electron fields, and 
they have a rather small number of unitary representations. These are
favourable features, in view of the stability criteria (S1), (S2) and
(S3) of Section \ref{3.3}.

\subsection{Virasoro minimal models}
\begin{figure}[!tbp]
\begin{center}
\input{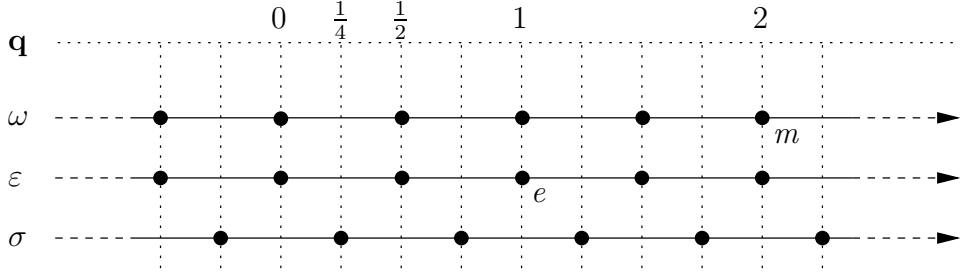}
  \parbox{14cm}{\caption{Physically realized representations for
      $\calc=\vir_1$; $e$\ marks the electron representation, $m$ a
      multi-electron \label{fig:repvir1}
representation and $\bullet$\ the representations that fulfill
\erf{eq:recipemonodromy2} and hence appear in the theory.}}
\end{center}
\end{figure}

The Virasoro minimal models are labelled by a strictly positive integer $k$.
They can be obtained from the coset construction, \cite{goko2}
\begin{equation}
  \vir_k \; \cong \; \frac{(\sutwo)_k\times(\sutwo)_1}{(\sutwo)_{k+1}}
  \quad, \qquad
  c_k = 1 -\frac{6}{(k+2)(k+3)}
  \quad.
\end{equation}
Each model has exactly one simple current $\vep$\ of order two
($\vep^{\ast2}=\om$), with vanishing self-monodromy $Q_{\vep}(\vep)=0$. Its
conformal weight is given by
\begin{eqnarray}
    \DE_{\vep}=\frac{k(k+1)}{4}=\frac{k}{4}\mod 1 \,,
   & \quad & \mathrm{ for } \, k\; \mathrm{ even} 
     \nonumber \\
  \DE_{\vep}=\frac{k(k+1)}{4}=-\frac{k+1}{4}\mod 1 \,,
  & \quad & \mathrm{ for } \, k\; \mathrm{ odd}  \, .
     \nonumber \\
\label{eq:sccw}
\end{eqnarray}
If we use these simple currents to construct the electron representation, then 
the charge-statistics connection \erf{eq:recipeweight} and relative locality
\erf{eq:recipemonodromy1} can be fulfilled if the
values of the Hall conductivity are restricted to 
\begin{eqnarray}
  k=1,2 \mod 4 & \Longrightarrow & \frac{1}{\sigma_H}=2,4,\ldots
      \nonumber \\
  k=3,4 \mod 4 & \Longrightarrow & \frac{1}{\sigma_H}=1,3,\ldots \quad.
\end{eqnarray}
These series of values of $\sigma_H^{-1}$\ can be obtained by applying the
shift map to the theories with $\sigma_H=1/2,1$, respectively.
We first consider the simplest example, $\calc=\vir_1$\ (Ising-model). The
largest possible value of the Hall conductivity is $\sigma_H=1/2$, which is an 
interesting Hall plateau,  since it is one of the few observed plateaux with 
an {\em even} denominator. 
The model has one nontrivial simple current, $\varepsilon$. The relevant
features of the model are summarized in Table \ref{tab:vir1}.
\begin{table}[!tbp]
\begin{center}
\begin{tabular}{|c||c||c|c|}
\hline
            & $\Delta$       & $\vep\ast(\cdot)$ & $Q_{\vep}(\cdot)$ \\ 
\hline\hline
$\Omega$    & 0              & $\vep$     & 0             \\ \hline
$\vep$      & $\frac{1}{2}$  & $\om$      & 0             \\ \hline
$\sigma$    & $\frac{1}{16}$ & $\si$      & $\frac{1}{2}$ \\ \hline
\end{tabular}
  \parbox{14cm}{\caption{Relevant features of the $\vir_1$-model: conformal
      weights of the representations,  the action of the simple current on the
  \label{tab:vir1}
      representations and the monodromy charges of the representations.}}
\end{center}
\end{table}
Applying the algorithm of Section \ref{sec:recipe}, we find a set of
representations, $\LA$, which is represented in Figure \ref{fig:repvir1}.
In this example, the Hall conductivity of the theory based on 
$(\sutwo)_1\times(\sutwo)_1$ which is described by a quantum Hall lattice,
and the Hall conductivity of the theory based on the coset 
$(\sutwo)_1\times(\sutwo)_1/(\sutwo)_2$ are identical (modulo shift map).
For a proof, see Appendix \ref{app:cosets}. %

The same procedure can be applied to the higher-$k$\ minimal models. The main
features of some of these theories are represented in Table \ref{tab:vir}.
We note that equation \erf{eq:sccw} together with 
the requirement $\DE_e\le7/2$ (see the remark concerning
the stability criterion (S2) in section \ref{3.3}) restricts the number of
minimal models that can be expected to describe a stable QHF by $k\le 3$. %
\begin{table}[!tbp]
\begin{center}
\begin{tabular}{|c|c||c|c|c||c|}
\hline
\multicolumn{2}{|c||}{Theory} & \multicolumn{3}{|c||}{Stability} & $\caln$ \\
\cline{1-5}
$\calc$ & $\sigma_H$ & $c$ & $\DE_e$ & $\nfra$ &  \\
\hline
$Vir_1$ & $\frac{1}{2}$ & $\frac{3}{2}$ & $\frac{3}{2}$ & 6 & 8\\
\hline
$Vir_2$ & $\frac{1}{2}$ & $\frac{17}{10}$ & $\frac{5}{2}$ & 12 & 8\\
\hline
$Vir_3$ & 1  & $\frac{9}{5}$ & $\frac{7}{2}$ & 10 & 4\\
\hline
\end{tabular}
  \parbox{14cm}{\caption{Features of theories constructed with
      Virasoro-minimal models in the neutral sector $\calc$. We indicate the
      central charge $c=c_\cala$, the electron conformal weight $\DE_e$\ and 
      the number of
      fractionally charged representations $\nfra$. In addition, we give the
      integer $\caln$\ that can be used to construct covariant characters under 
\label{tab:vir}
      $\GA_2(S)$\ (see Appendix \ref{app:modular}).}} 
\end{center}
\end{table}
\clearpage

\subsection{Simple current extensions of Virasoro minimal models}

The (unique) simple current of a $\vir_k$-model has integer conformal weight
for $k=3,4\mod 4$, in which case the model allows for an extension of the
chiral algebra by the primary field corresponding to the simple current. It
is then possible that in the set of representations of the {\em extended} 
algebra some new simple currents are encountered. 

\begin{figure}[!tbp]
\begin{center}
\input{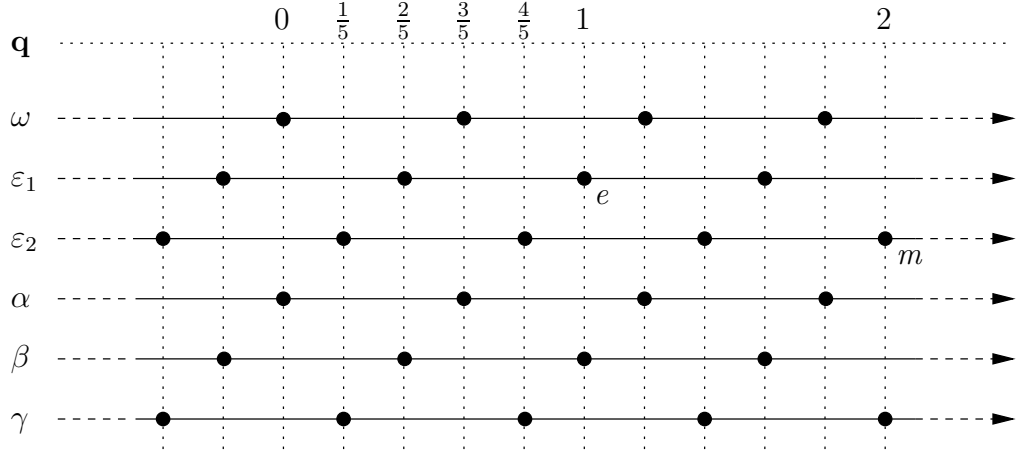}
  \parbox{14cm}{\caption{Representations for $\calc=W_3$; $e$\ marks the electron representation, $m$ a multi-electron
  \label{fig:repw3}
representation and $\bullet$\ the representations that fulfill
\erf{eq:recipemonodromy2} and hence can appear in the theory.}}
\end{center}
\end{figure}

We first consider the easiest example, with $k=3$. The simple current of
$\vir_3$\ has conformal weight $\DE=3$, and the extension is known as the
$W_3$-minimal model or three-states Potts-model. Its simple current group is
$\{\om,\vep,\vep'\}\cong\zet_3$. The relevant
features of the model are given in Table \ref{tab:potts}.
\begin{table}[!tbp]
\begin{center}
\begin{tabular}{|c||c||c|c||c|c|}
\hline
               & $\DE       $   & $\vep\ast$  & $Q_{\vep}(.)$ & 
                                  $\vep'\ast$ & $Q_{\vep'}(.)$ \\ 
\hline\hline
$\Omega$   & 0              & $\vep$  & 0             & 
                              $\vep'$ & 0    \\ 
\hline
$\vep$     & $\frac{2}{3}$  & $\vep'$ & $\frac{2}{3}$ & 
                              $\om$   & $\frac{1}{3}$ \\ 
\hline
$\vep'$    & $\frac{2}{3}$  & $\om$   & $\frac{1}{3}$ & 
                              $\vep$  & $\frac{2}{3}$ \\ 
\hline
$\al$      & $\frac{2}{5}$  & $\be$   & 0             & 
                              $\ga$   & 0             \\ 
\hline
$\be$      & $\frac{1}{15}$ & $\ga$   & $\frac{2}{3}$ & 
                              $\al$   & $\frac{1}{3}$ \\ 
\hline
$\ga$      & $\frac{1}{15}$ & $\al$   & $\frac{1}{3}$ & 
                              $\be$   & $\frac{2}{3}$ \\ 
\hline
\end{tabular}
  \parbox{14cm}{\caption{Relevant features of the $W_3$-model: conformal
      weights of the representations,  the action of the simple current on the
  \label{tab:potts}
      representations and the monodromy charges of the representations.}} 
\end{center}
\end{table}
We may use a simple current, $\vep$, say, to construct the electron
representation, but {\em not} both simple currents $\vep$\ and $\vep'$, 
because this would violate the relative locality requirement. Equations
\erf{eq:recipeweight} (charge-statistics connection) and 
\erf{eq:recipemonodromy1} (relative locality) can be fulfilled if the
Hall conductivity is given by
\begin{equation}
  \frac{1}{\si_H}=\frac{5}{3}+2l
\end{equation}
with $l$\ a positive integer. 
Applying the construction of Section \ref{sec:recipe} we find a set of
representations, $\LA$, represented in Figure \ref{fig:repw3}.

Let us finally consider the case $k=4$. The simple current has conformal
weight $\DE=5$, and the extension is known as the $W_5$-minimal model. It 
has nine
unitary representations and a $\zet_3$-group $\{\om,\vep,\vep'\}$\ of simple
currents, with $\DE_{\vep}=\DE_{\vep'}=\frac{4}{3}$. The construction of Section
\ref{sec:recipe} can be applied using $\vep$\ to construct the electron
representation; the values of the Hall conductivity are restricted to
$1/\si_H=\frac{1}{3}+2l$, with $l$\ a positive integer.
The main
features of the above theories are represented in Table \ref{tab:scext}.
\begin{table}[!tbp]
\begin{center}
\begin{tabular}{|c|c||c|c|c||c|}
\hline
\multicolumn{2}{|c||}{Theory} & \multicolumn{3}{|c||}{Stability} & $\caln$\\
\cline{1-5}
$\calc$ & $\sigma_H$    & $c$            & $\DE_e$       & $\nfra$ & \\
\hline
$W_3$   & $\frac{3}{5}$ & $\frac{9}{5}$  & $\frac{3}{2}$ & 10 & 60\\
\hline
$W_5$   & $3$           & $\frac{13}{7}$ & $\frac{3}{2}$ &  3  & 12\\
\cline{2-6}
        & $\frac{3}{7}$ & $\frac{13}{7}$ & $\frac{5}{2}$ &  21 & 84\\
\hline
\end{tabular}
  \parbox{14cm}{\caption{Features of theories constructed with
      $W_3$- and $W_5$-models in the neutral sector $\calc$. We indicate the
      central charge $c=c_\cala$, the electron conformal weight $\DE_e$\ 
      and the number of
  \label{tab:scext}
      fractionally charged representations $\nfra$. In addition, we give the
      integer $\caln$\ that can be used to construct covariant characters
      under $\GA_2(S)$\ (see Appendix \ref{app:modular}).}} 
\end{center}
\end{table}


\begin{appendix}
\section{Modular covariance under $\GA_2(S)$}
\label{app:modular}

Suppose, for simplicity, that there is only one species of electrons.
Then we show that the linear space spanned by certain characters of the 
theories constructed according to the algorithm of Section \ref{sec:recipe}
carries a representation of a subgroup of
the modular group $SL(2,\zet)$, usually denoted by $\GA_2(S)$. This group is 
generated by the operators $S$\ and $T^2$ introduced in Section \ref{3.2}.

The proof of our claim goes as follows.

\begin{itemize}

\item[(i)] The even integer $\caln$\ can be chosen such that, for a suitably
  chosen positive integer $n$, the $n$-fold fusion of
  the electron representation is given by
\begin{equation}
  e^{\ast n}\;=\;(\vep^{\ast n},\frac{n}{\sqrt{\si_H}})\;=\;(\om,\sqrt{\caln})
  \,,
\end{equation}
where the electron representation is given by 
  $e=(\vep,\frac{d}{\sqrt{\caln}})$, with $d$\ a positive integer. We have 
  that $\caln=nd$.

\item[(ii)] The representations of $\uone$-current algebra labelled by the
  set $\{r=\frac{l}{\sqrt{\caln}}|l\in\zet\}$\ can be grouped
  into $\caln$\ (reducible) $\uone$-representations by building the direct sum of
  representation spaces 
\begin{equation}
  \bar{\calh}_k\;=\;
  \bigoplus_{l\in\zet}\calh_{(\frac{k+\caln l}{\sqrt{\caln}})}\;,
  \qquad 0\leq k \leq \caln-1
  \quad.
\end{equation}
The linear space spanned by the corresponding characters 
\begin{equation}
  \bar{\chi}_k(\tau)=\tr_{\bar{\calh}_k}[e^{2\pi i\tau(L_0-\frac{c}{24})}]
\end{equation}
carries a representation of the modular group given by the matrices
\begin{equation}
  \bar{T}_{kl}  =  \de_{kl}e^{2\pi i\DE_k},\qquad \DE_k=\frac{k^2}{2\caln} 
  \quad,
\end{equation}
and
\begin{equation}
  \bar{S}_{kl}  =  \frac{1}{\sqrt\caln}e^{-2\pi i(\frac{kl}{\caln})}
  \quad.
\end{equation}

\item[(iii)] The space of characters of the $\calc$-theory, defined by
\begin{equation}
  \chi_{\pi}(\tau)=\tr_{\calh_{\pi}}[e^{2\pi i\tau(L_0-\frac{c}{24})}]
  \quad,
\end{equation}
carries a representation of the modular group, given by
the diagonal matrix $T_{\pi\pi'}=\de_{\pi\pi'}e^{2\pi i(\DE_{\pi}-c/24)}$\ and
some unitary, symmetric matrix $S_{\pi\pi'}$.

\item[(iv)] The (reducible) representations of the algebra $\calc\times\uone$
on the spaces
\begin{equation}
  \calh_{(\pi,k)}=\calh_{\pi}\otimes \bar{\calh}_k
\end{equation}
have characters 
\begin{equation}
  \bar{\chi}_{(\pi,k)}(\tau)=\chi_{\pi}(\tau)\cdot\bar{\chi}_k(\tau)
  \quad.
\end{equation}
These characters are considered as linearly independent 
and span a finite dimensional vector space, which we denote as
  $\mathrm{Char}_{(\PI,\caln)}$. 
It carries a representation of the modular group, given by matrices
$\bar{T}_{(\pi,k)(\pi',k')}=T_{\pi\pi'}\cdot \bar{T}_{kk'}$\ and $\bar{S}_{(\pi,k)(\pi',k')}=S_{\pi\pi'}\cdot \bar{S}_{kk'}$. 

\item[(v)] The set of representations
  $\{(\pi,k)|\pi\in\PI,k=0,\ldots,\caln-1\}$\ decomposes into orbits under the 
  action
  of the abelian group of simple currents $\{\om,e,\ldots,e^{\ast (n-1) }\}$, the
  action being given by  
\begin{equation}
  e\ast(\pi,k)=(\vep\ast\pi,k+d\mod\caln)
  \quad.
\end{equation}
The orbits have all the same length, $n$.

\item[(vi)] Let us define characters of the orbits by setting
\begin{equation}
  \label{eq:orbitcharacters}
  \hat{\chi}_{\lfloor\pi,k\rfloor}=
     \bar{\chi}_{(\pi,k)}+\bar{\chi}_{e\ast(\pi,k)}+\ldots+\bar{\chi}_{e^{\ast n}\ast(\pi,k)} \,,
           \qquad 0\leq k \leq d-1
  \quad.
\end{equation}
Consider only those characters $\hat{\chi}$, for which
$(\pi,\frac{k}{\sqrt{\caln}})$\ is in the set of physically realizable
  representations $\LA$. They are linearly independent and span a subspace of
 $\mathrm{Char}_{(\PI,\caln)}$, which we denote as
 $\mathrm{Char}_{\LA}$. These characters involve exactly those representations 
 that appear in the Quantum Hall CCFT.

\item[(vii)] We now prove that the subspace $\mathrm{Char}_{\LA}$\ of
  $\mathrm{Char}_{(\PI,\caln)}$\ is invariant under the action of $T^2$\ and
  $S$. 
  For $T^2$, this follows from the fact that, in an orbit, there appear
  only representations whose conformal weights differ by half-integers.
  For, we have that $\DE_e$\ is half-integer and that $Q_e(\la)=0$, and
  therefore the difference $\DE_{e\ast\la}-\DE_{\la}=\DE_e$\ is a
  half-integer. Thus, we have that
\begin{equation}
  (T^2\hat{\chi}_{\lfloor\pi,k\rfloor})(\tau)\;=\;
  \hat{\chi}_{\lfloor\pi,k\rfloor}(\tau+2)\;=\;
  e^{2\pi i\cdot (2\DE_{(\pi,k)}-c/12)}\hat{\chi}_{\lfloor\pi,k\rfloor}(\tau)
  \quad.
\end{equation}
For $S$, we reason as follows. Let $\LA$\ be a finite set of
representations of a chiral algebra for which there is a unitary and symmetric 
matrix $S$\ implementing the transformation $\tau\longmapsto-1/\tau$\ on
the characters. Let $\la,\la'$\ be representations and $e$\ a simple current
in $\LA$; then we have that $S_{e\ast\la,\la'}=S_{\la,\la'}e^{2\pi i\cdot
  Q_e(\la')}$. A simple calculation shows that
\begin{equation}
  (S\hat{\chi}_{\lfloor\pi,k\rfloor})(\tau)\;=\;
  \hat{\chi}_{\lfloor\pi,k\rfloor}(-\frac{1}{\tau})\;=\;
  \sum_{(\pi',k')} nS_{(\pi,k)(\pi'k')}\;\hat{\chi}_{\lfloor\pi',k'\rfloor}(\tau)
  \quad.
\end{equation}
Thus, we have proven that the space of characters $\mathrm{Char}_{\LA}$\
carries  a representation of the group $\GA_2(S)$, and that, in the canonical
basis \erf{eq:orbitcharacters}, the corresponding matrices read
\begin{equation}
  \hat{T}^2_{\lfloor\pi,k\rfloor,\lfloor\pi',k'\rfloor}\;=\;
      \de_{\lfloor\pi,k\rfloor,\lfloor\pi',k'\rfloor}
     e^{2\pi i\cdot (2\DE_{(\pi,k)}-c/12)}
  \quad,
\end{equation}
and
\begin{equation}
  \hat{S}_{\lfloor\pi,k\rfloor,\lfloor\pi',k'\rfloor}\;=\;
         n\bar{S}_{(\pi,k),(\pi',k')}
  \quad.
\end{equation}
Since $\bar{S}$\ is symmetric, the same is true for $\hat{S}$; $\hat{S}$\ is
also unitary, since it is the restriction of a unitary linear map on an
invariant subspace.

\end{itemize}

\section{Cosets}
  \label{app:cosets}

The coset construction \cite{baha,goko} provides a powerful tool
to construct (chiral) conformal field theories. In this appendix,
we describe the main idea and sketch a few important features of this
construction.

The starting point is a WZW theory, i.e., a theory based on
non-abelian currents. The zero-modes of these currents form a Lie
algebra $\g$. Natural examples of particular interest for our purposes are
provided by WZW theories based on simply laced Lie algebras at level
one: they can also be described by a lattice theory based on the root
lattice of the Lie algebra. The affine Sugawara construction provides
a chiral stress energy tensor, $T(z) = \sum_{n\in\zet} L_n z^{-n-2}$,
whose Fourier modes $L_n$ span a Virasoro algebra with a certain central 
charge $c$.
(In the case of the lattice model, $c$ equals the rank of the lattice.)

Next, one fixes a subalgebra, $\g'$, of $\g$. The affine Sugawara construction
applied to $\g'$ yields another Virasoro algebra $L_n'$, with a different 
central charge $c'\leq c$. The crucial observation is that the operators
\begin{equation}
          \dot L_n := L_n - L_n' 
\end{equation}
form a third Virasoro algebra, with central charge $\dot c=c-c'$ in the
range $c > \dot c \geq 0$, which is the Virasoro algebra of the so-called
coset theory. The coset construction therefore lowers the
Virasoro central charge.

The conformal weights $\dot\Delta$ in the coset theory are given
by differences of the conformal weights,
\begin{equation}
\dot\Delta = \Delta-\Delta'  \bmod \zet
\end{equation}
In particular, the currents in $\g'$ have zero conformal weight:
they are `gauged'. Indeed, full coset CFT's admit a description as gauged
WZW theories \cite{fegk2}.

As a consequence, the starting point for the construction of the state space
of a coset theory are so-called branching spaces, i.e., the spaces of
multiplicities of $\g'$-representations in $\g$-representations.
The precise construction of the state space is actually quite subtle, and
we refer the reader to \cite{scya5,fusS4} for details. In particular,
despite numerous claims in the literature, the spaces of physical states are in
general {\em not} the branching spaces, but suitable subspaces thereof.

We only mention
that, in analogy to the $\zet_2$ symmetry of the Kac table, typically
different branching spaces provide different representatives for one and
the same physical state. This effect, which goes under the name of
``field identification'', can be understood in almost all cases in terms
of group theoretical selection rules. Additional subtleties occur if 
this field identification has so-called fixed points. In this case, branching
spaces have to be split and one branching space contains states of different
primary fields. The corresponding algorithm has been worked out for
diagonal cosets in \cite{fusS4}.

As an illustration of the concepts involved in the coset construction, we
provide a criterion for those cases in which the theory based on the coset
describes a QHF with the same Hall conductivity as the original theory.
If $\epsilon$ is a simple current of the WZW theory based on the Lie algebra
$\g$ with conformal weight $\DE_{\epsilon}$ and if $\lambda'$ is a simple
current of the theory based on the subalgebra $\g'$, with conformal weight
$\DE_{\lambda'}$, such that the branching space associated to $\epsilon$ and
  $\lambda'$ is non trivial, then there is a simple current
  $(\epsilon,\lambda')$ of the coset theory with conformal weight 
\begin{equation}
\DE_{(\epsilon,\lambda')} = \DE_{\epsilon} -\DE_{\lambda'} \;\mod \zet
\end{equation}
In particular, if $\lambda'$ is the vacuum representation of the $\g'$ theory, 
then $\DE_{(\epsilon,\lambda')} = \DE_{\epsilon}\;\mod\zet$. Thus if $\epsilon$
is the electrically neutral part of a one-electron representation of the
$\uone\otimes\g$-theory, and $\lambda'$ is the vacuum representation of the
$\g'$ theory, then $(\epsilon,\lambda')$ is the electrically neutral part
of a one-electron representation of the coset theory. In this case, formula
\erf{eq:recipeweight}, i.e., 
\begin{equation}
\frac{1}{\sigma_H} +2\Delta_{\epsilon} = 
\frac{1}{\dot\sigma_H} + 2\Delta_{(\epsilon,\lambda')} = 1 \;\mod 2
\end{equation}
tells us that the original theory and the coset construction have the {\em
same} Hall conductivity, modulo the shift map. 
In particular, this criterion can be used to
identify possible coset constructions based on maximally symmetric quantum
Hall lattices.

\end{appendix}

 \vskip3em
\small
 \newcommand\wb{\,\linebreak[0]} \def\wB {$\,$\wb}
 \newcommand\Bi[1]    {\bibitem{#1}}
 \newcommand\J[5]   {{\sl #5}, {#1} {#2} ({#3}) {#4} }
 \newcommand\PhD[2]   {{\sl #2}, Ph.D.\ thesis (#1)}
 \newcommand\Prep[2]  {{\sl #2}, preprint {#1}}
 \newcommand\BOOK[4]  {{\em #1\/} ({#2}, {#3} {#4})}
 \def\jf    {J.\ Fuchs}
 \newcommand\inBO[7]  {in:\ {\em #1}, {#2}\ ({#3}, {#4} {#5}), p.\ {#6}}
 \newcommand\inBOx[7] {in:\ {\em #1}, {#2}\ ({#3}, {#4} {#5})}
 \newcommand\gxxI[2] {\inBO{GROUP21 Physical Applications and Mathematical
              Aspects of Geometry, Groups, and \A s{\rm, Vol.\,2}}
              {H.-D.\ Doebner, W.\ Scherer, and C.\ Schulte, eds.}
              \WS\Si{1997} {{#1}}{{#2}}}
 \def\ahp   {Ann.\wb Henri Poincar\'e}
 \def\anop  {Ann.\wb Phys.}
 \def\baps  {Bull.\wb Am.\wb Phys.\wb Soc.}
 \def\comp  {Com\-mun.\wb Math.\wb Phys.}
 \def\duke  {Duke\wB Math.\wb J.}
 \def\duki  {Duke\wB Math.\wb J.\ (Int.\wb Math.\wb Res.\wb Notes)}
 \def\foph  {Fortschr.\wb Phys.}
 \def\ijmp  {Int.\wb J.\wb Mod.\wb Phys.\ A}
 \def\ijmb  {Int.\wb J.\wb Mod.\wb Phys.\ B}
 \def\imrn  {Int.\wb Math.\wb Res.\wb Notices}
 \def\jams  {J.\wb Amer.\wb Math.\wb Soc.}
 \def\jetl  {JETP letters}
 \def\jopa  {J.\wb Phys.\ A}
 \def\jstp  {J.\wb Stat.\wb Phys.}
 \def\jhep  {J.\wb High\wB Energy\wB Phys.}
 \def\mape  {Math.\wb Phys.\wb Electr.\wb J.}
 \def\mpla  {Mod.\wb Phys.\wb Lett.\ A}
 \def\natu  {Nature}
 \def\nupb  {Nucl.\wb Phys.\ B}
 \def\phlb  {Phys.\wb Lett.\ B}
 \def\phma  {Phil.\wb Mag.}
 \def\phrb  {Phys.\wb Rev.\ B}
 \def\phrd  {Phys.\wb Rev.\ D}
 \def\phrl  {Phys.\wb Rev.\wb Lett.}
 \def\remp  {Rev.\wb Mod.\wb Phys.} 
 \def\susc  {Surf.\wb Sci.}
 \def\thmp  {Theor.\wb Math.\wb Phys.}
 \newcommand\geap[2] {\inBO{Physics and Geometry} {J.E.\ Andersen, H.\
            Pedersen, and A.\ Swann, eds.} \MD\NY{1997} {{#1}}{{#2}} }
 \def\AMS    {{American Mathematical Society}}
 \def\AP     {{Academic Press}}
 \def\AW     {{Addison-Wesley}}
 \def\CUP    {{Cambridge University Press}}
 \def\LMS    {{London Mathematical Society}}
 \def\MD     {{Marcel Dekker}}
 \def\NH     {{North Holland Publishing Company}}
 \def\SV     {{Sprin\-ger Ver\-lag}}
 \def\TE     {{Teubner}}
 \def\WS     {{World Scientific}}
 \def\Ad     {{Amsterdam}}
 \def\Be     {{Berlin}}
 \def\Ca     {{Cambridge}}
 \def\Le     {{Leipzig}}
 \def\NY     {{New York}}
 \def\PR     {{Providence}}
 \def\RC     {{Redwood City}}
 \def\Si     {{Singapore}}


\end{document}